\documentclass[preprint,review,12pt]{elsarticle}

\usepackage{graphicx}
\usepackage{amssymb}
\usepackage{color}
\usepackage{lineno}
\usepackage{textgreek}
\usepackage{array}
\usepackage{tabularx}
\usepackage{lscape}
\usepackage{multirow}
\usepackage{footnote}
\usepackage[table]{xcolor}
\usepackage[font={scriptsize,it}]{caption}
\usepackage{paralist}
\usepackage{sidecap} 
\usepackage{hyperref}
\hypersetup{
    colorlinks=true,
    linkcolor=blue,
    urlcolor=blue,
}

\usepackage{natbib}
\setcitestyle{authoryear,open={(},close={)}}

\journal{Icarus}

\begin{document}

\begin{frontmatter}


\title{Reflectance study of ice and Mars soil simulant associations -- I. H$_2$O ice.}

\author{Zuri\~ne Yoldi\textsuperscript{1}*}
\author{Antoine Pommerol\textsuperscript{1}}
\author{Olivier Poch\textsuperscript{1,2}}
\author{Nicolas Thomas\textsuperscript{1}}

\address{\textsuperscript{1}Physikalisches Institut, Universit\"at Bern and NCCR PlanetS, Sidlerstrasse 5, 3012 Bern, Switzerland.}
\address{\textsuperscript{2}Univ. Grenoble Alpes, CNRS, IPAG, 38000 Grenoble, France}
\address{\textsuperscript{*} Now at the section for the Physics of Ice, Climate and Earth at the Niels Bohr Institute, University of Copenhagen.}
\tnotetext[label0]{Formal publication in Icarus:}
\tnotetext[label1]{\url{https://doi.org/10.1016/j.icarus.2020.114169}}

\begin{abstract}
The reflectance of water ice and dust mixtures depends, amongst other parameters, on how the components are mixed (e.g. intimate mixture, areal mixture or coating) \citep{Clark_1999}. Therefore, when inverting the reflectance spectra measured from planetary surfaces to derive the amount of water ice present at the surface, it is critical to distinguish between different mixing modes of ice and dust. However, the distinction between mixing modes from reflectance spectra remains ambiguous. Here we show how to identify some water ice/soil mixing modes from the study of defined spectral criteria and colour analysis of laboratory mixtures. We have recreated ice and dust mixtures and found that the appearance of frost on a surface increases its reflectance and flattens its spectral slopes, whereas the increasing presence of water ice in intimate mixtures mainly impacts the absorption bands. In particular, we provide laboratory data and a spectral analysis to help interpret ice and soil reflectance spectra from the Martian surface. 
\end{abstract}


\end{frontmatter}

%%\linenumbers

\section{Introduction}
\label{S:1}

Water has played a key role in the early evolution of Mars. It has shaped its surface at various scales, carving valley networks terminated by deltas, and producing large outflow channels leading to the shorelines of a putative ocean that may have covered a significant fraction of the northern hemisphere \citep{carr2010}. Liquid water has also altered the mineralogical composition of the primordial mafic crust of Mars over depths of kilometres \citep{bibring2006, ehlmann2011} and might have made and maintained Mars habitable during part of its first billion years \citep{vago2017}. In the current cold and dry Amazonian climate, water is only present at the surface and in the close subsurface in the form of ice and as a trace gas in the atmosphere \citep{Lasue2013}. The evidence for liquid water (Recurring Slope Lineae) remains controversial \citep{McEwen:2014, Ojha:2015, Dundas:2017}. Ice is nevertheless still thought to play an active role in the current Amazonian area, the ice being regularly redistributed between various reservoirs following astronomical forcing of the climate \citep{Head2003}.

The largest known reservoirs of ice are the permanent polar caps which are kilometres-thick layered deposits of ice mixed with dust in various proportions that cover each Martian pole \citep{smith2020}. A major difference between the two polar caps is that the northern polar cap is exposed whereas the southern polar cap is covered by a layer of dust that masks, almost entirely, the water ice \citep{bibring2004, Plaut2007}. A small permanent cap of CO$_2$ ice is present directly above the south pole \citep{thomas2020}. The exposed water ice of the Northern permanent polar cap is the main source of water vapor which has been found in highly variable, trace amounts in the martian atmosphere \citep{smith2002}. The active seasonal cycle characterised by the extension and recession of the metre-thick seasonal polar caps, down to latitudes of $\sim$50$^{\circ}$, is dominated by the condensation/sublimation equilibrium between atmospheric CO$_2$ and solid CO$_2$ ice at the surface \citep{Leighton:1966}. Water ice is only present in low amounts in the northern hemisphere and at a few locations in the southern hemisphere but its presence still influences strongly the recession of the cap in Spring \citep{appere2011, Pommerol2013}.

In addition to the yearly seasonal cycle, the accumulation and sublimation of surface ice responds to longer term climatic cycles controlled by the periodic and secular variations of astronomical parameters \citep{laskar2004}. In particular, large excursions of the obliquity are responsible for a redistribution of the ice between the thick, permanent, polar caps observed today and metre-thick accumulations of dusty ice over significant fractions of mid- and high-latitudes in both hemispheres \citep{Head2003, levrard2004}. Various morphological features provide evidence of recent redistribution of water ice to mid-latitudes. These include near-surface ground ice \citep{mustard2001}, gullies \citep{conway2019}, polygons and other patterned ground \citep{mangold2004}, viscous flow features \citep{milliken2003}, and rocky glaciers \citep{Head2003}. Water ice in the close subsurface below a thin desiccated layer of dust is found at latitudes poleward of 60$^{\circ}$ in both hemispheres with a concentration often exceeding 50 wt$\%$ \citep{feldman2008}.

While geophysical techniques such as neutron spectroscopy \citep{feldman2008} and ground penetrating radar \citep{Plaut2007} are used to detect and characterize ice in the subsurface, most of our knowledge about surface ice is gained by optical remote-sensing. High-resolution visible imaging and infrared spectroscopy and spectral imaging are techniques of choice to identify water ice at the surface and decipher some of its properties both from orbit and in-situ. Visible imaging provides the highest spatial resolution and the contrast between bright ice and the dark surface at the shortest wavelength makes it particularly sensitive to ice, even in low amounts. As the seasonal cycle of volatiles on Mars is dominated by CO$_2$, not H$_2$O, the infrared spectral range is particularly useful to distinguish H$_2$O from CO$_2$ from their distinctive spectral features \citep{Langevin2007}. 

Recent examples of detection of ice reservoirs by optical remote-sensing on Mars include orbital observations of ground ice at mid-latitudes revealed by recent impact craters \citep{Byrne:2009} and pits with scarps exposing ice near 55$^{\circ}$N latitude \citep{Dundas:2018}. In both cases, the presence of ice was inferred from high-albedo bluish patches on high-resolution visible images and confirmed by the observation of diagnostic absorption bands of solid H$_2$O at 1.5 and 2 \textmu m. The presence of ice at these latitudes could be the result of the deposition of ice-rich mantling units, possibly as snowfall, in conditions of higher obliquity.

Water ice has also been observed at the surface by two landed missions: Viking 2 in 1976 \citep{Jones:1976} and Phoenix in 2008 \citep{Mellon:2009}. At 48$^{\circ}$N latitude, the Viking 2 camera observed diurnal frost deposited at night that rapidly sublimed after sunrise. Although the nature of the frost (H$_2$O or CO$_2$) had initially been debated, surface energy balance calculations show that this behaviour is only consistent with H$_2$O and not CO$_2$ \citep{hart1986, svitek1990}. At 68$^{\circ}$N latitude, Phoenix used a robotic scoop to excavate up to 20 cm of dust, revealing below the ice-rich ground first detected from orbit by neutrons spectroscopy. During late summer, both snowfall and surface frost were also observed \citep{Smith:2009}. 

Beyond the detection and identification of ices, either through reflectance contrast or distinctive spectral fingerprints, optical reflectance data should also provide information on various properties of the ice itself and its mode of association to other constituents of the surface. Mars is a dusty planet and the abundant red oxidized dust transported by the winds contaminates at various levels all occurrences of ice found at the surface and, over history, has led to contaminated sub-surface layers. Variable levels of contamination by dust are for instance responsible for the layering seen in the polar layered deposits of the permanent polar caps. The variable dust-to-ice ratios are interpreted in terms of sublimation-condensation equilibrium and hence past insulation history at the pole. The dust-to-ice ratio in the Martian subsurface is also key to understanding the origin and emplacement mechanism of the ice. If the amount of ice is below the open porosity of the regolith, the ice can have been emplaced by diffusion of water from the atmosphere. If the concentration of ice is found to be in excess of the porosity however, it requires a different deposition mechanism such as a snowfall, which has strong implication for past climate.

The scattering of light by ice and its contaminants is the fundamental physical process that permits all identification and characterisation by optical remote-sensing. In all situations investigated in this work, binary mixtures and associations between H$_2$O ice and dust are considered. If such a system might seem simple at first glance, a huge level of complexity resides in the multiple ways the ice and the dust can be associated and the resulting multiple scattering by the two types of materials. A well-known effect of mixing ice with a fine-grained dark contaminant is the strong darkening of the sample, out of proportion with the concentration of the dark contaminant \citep{Yoldi2015}. This is well studied on the Earth where minute amounts of fine dark soot particles are known to be able to darken snow considerably \citep{Yasunari2011, He2019}.

Despite constant progress over the past decades, physical modeling of the light scattering process remains difficult, both because of challenges associated with the physics and its mathematical implementation and because of the difficulty in describing and representing the complexity of natural samples. Even for a binary system with water ice as one component and mineral dust as the other component, calculating the reflectance of the mixture requires detailed knowledge of various properties of the particles which are often out of reach in the case of planetary remote-sensing. Laboratory experiments with well-controlled samples are therefore of considerable value in testing models and results can also be used for direct comparison with planetary data. 

While spectro-photometric measurements with refractory materials such as rock or mineral powders abound in the literature and in open databases, good-quality measurements with icy samples are more difficult to find. The reason is not merely the need for the measurement equipment to work at low temperature but rather the challenge to produce the ice-bearing samples in a reproducible way and characterize them thoroughly. While it is possible to gain additional understanding of the scattering process by using stable analogues for ice such as glass beads or other transparent particles \citep{Jost2013}, the unique properties of the water molecule such as high surface tension, capillarity... make the interactions between ice and dust unique and experimentation with actual material necessary.  

\citet{Clark:1981} has measured the reflectance of red cinder from Mauna Kea spread over water frost as well as frost grown on the same red cinder and charcoal. Strong differences between the behaviour at visible and near-infrared wavelengths are visible as the effect of frost growth is more obvious at the shortest wavelengths where the reflectance of the cinder is the lowest. \citet{hart1986} have used these experimental spectra to estimate the thickness of frost seen by Viking 2 to 50 \textmu m on the basis of the observed relative increase of reflectance at 0.8 \textmu m. Unfortunately, spectral data were restricted to wavelengths greater than 0.65 \textmu m and the blue wavelengths where the contrast of reflectance is the highest was not covered. In the continuum of the near-infrared, little change is visible for the red cinder, which is very bright but strong differences are visible for the charcoal, which develops a blue slope over the entire spectrum. The asymmetric absorption bands of water-of-hydration at 1.5 and 2.0 \textmu m progressively become more symmetric and shift their centres towards longer wavelengths. These data are very useful to understand the spectral effects of frost growth over two very different substrates but these substrates are relatively poor analogues of actual Martian surfaces. Indeed, the Mauna Kea red cinder is brighter than the brightest Martian surfaces by a factor of two and the charcoal is darker than the darkest Martian areas by a factor of two and lacks the prominent visible red slope seen everywhere on Mars.

\citet{Clark:1981} and \citet{Clark_1984} also produced mixtures of the same dust analogue from the Mauna Kea as well as a clay mineral (kaolinite) and charcoal with water ice. These mixtures were produced by fast freezing a suspension of dust particles into liquid water. \citet{roush1990} modified the sample preparation protocol by grinding and sieving the dusty ice produced to isolate particles with a diameter of about 100 \textmu m. The spectral data were extended to the mid-infrared but lack the visible range ($<$ 0.65 \textmu m).

\citet{Pommerol2013b} have measured the bidirectional visible reflectance of a few plausible analogues for Martian icy surfaces, including frost and frozen soils. They found that, beyond differences in overall albedo, the different types of associations between ice and mineral dust result in totally different angular dependencies of the reflectance. 

\citet{Yoldi2015} have shown that two models used in planetary sciences, \citep{Hapke1993, Hiroi1994}, can accurately predict the reflectance of a mixture of spherical water ice particles and irregular basalt particles, but only if the properties of the particles are known, including some parametrization of their complex shape. This study was performed at a single wavelength however and spectral or colour effects were not studied.

\citet{Kaufmann:2015} studied the penetration of solar radiation into both pure and dust-contaminated snow and pointed out the major role played by small-scale inhomogeneities in the way light penetrates an ice-dust mixture.

Recently, \citet{gyalay2019} have measured the reflectance of intimate mixtures of water ice and JSC Mars-1 in the visible spectral range (400-1000 nm) and use the results to estimate the amount of water ice in the Martian soils from the images taken in-situ by the multispectral camera of the Phoenix lander. They reached conclusions about the presence of both pore-filling ice which diffused into the regolith from the atmosphere and ice in excess of the porosity which could have evolved from past ice-rich deposits.

Here, we present measured reflectance spectra of different types of associations between water ice and dust analogues that are relevant for the different ways in which ice occurs, that have been presented briefly above. Binary intimate mixtures (i.e. mixtures at the scale of the individual grains) of granular ice and dust are first-order analogues for the dusty granular ice which can be found at the surface of the permanent Northern polar cap (the water-rich annulus around the northern seasonal cap or exposed mid-latitude icy deposits). We produce these well-characterized samples using the equipment and protocols developed over a decade at the The Laboratory for Outflow Studies of Sublimating icy materials (or LOSSy) at the University of Bern \citep{Pommerol:2019}. Here, our approach is similar to the one followed by \citet{Clark:1981} and \citet{Clark_1984} but our mixtures are actual intimate particulate mixtures of ice and dust and strongly differ from the frozen suspensions used by \citet{Clark:1981}. Our intimate mixtures resemble closely the ones of \citet{gyalay2019} with the difference that they use grinded and sieved ice particles, probably irregular, while we use spherical particles.

In addition to these well-characterized binary mixtures which are ideal for comparisons with physical models, we produce two additional types of associations between water ice and dust analogues: frost growing over the mineral substrates and dust grains embedded into a matrix of ice. We simulate the process of direct frost growth over cold mineral substrates as analogues to the diurnal and seasonal frosting occur on Mars. This approach is again relatively similar to the one adopted by \citet{Clark:1981} but we extend the spectral range to shorter visible wavelengths where the contrast is the highest and where visible colour cameras can detect the smallest amounts of frost. We also produce icy soils with their porosity saturated by ice by freezing wet soils. Although the production process itself is not representative of any contemporary Martian surface process, the structure of the resulting samples is probably representative of small Martian icy soils where the entire porosity has been filled by ice.

Two types of dust analogues are chosen which represent the two end-members of surfaces found on Mars: bright and red soils and dark basaltic terrains. For the bright red soils, the widely used JSC Mars-1 Hawaiian palagonite \citep{Allen:1997} was used. For dark basaltic terrains, powder produced from Hawaiian basalts identical to the one used by \citet{Pommerol2013b} was used.

A versatile VIS-NIR (0.4 - 2.4 \textmu m) hyperspectral imaging system allows us to produce spectral and imaging data that mimic the ones acquired on Mars by different colour imagers and imaging spectrometers. Hence, we can assess the usefulness of various spectral/colour criteria in both spectral ranges and explore the complementarity between the two techniques. These spectral data are useful for the direct interpretation of optical remote-sensing of Mars as well as the calibration and verification of physical models. In addition to presenting the samples and the spectral data that are also made available in open databases, we study the behaviour of a range of spectral criteria and discuss how the different types of associations between ice and dust can be distinguished and characterized using reflectance spectroscopy and colour imaging.

\section{Setup, material and data analysis}
\label{S:3}

\subsection{Instrument description}
\label{S:3.2}
All the measurements were made in the Simulation Chamber for Imaging the Temporal Evolution of Analogue Samples (SCITEAS), at LOSSy. SCITEAS is a thermal vacuum chamber in which large samples are exposed to low pressure and temperature conditions (secondary vacuum of 10$^{-6}$ mbar and samples kept at about 220 K) if desired. The chamber is equipped with a large quartz window at the centre of its upper lid, which is used for optical observations of the surface of the sample. Here, we have used a hyperspectral imaging system to characterize the surfaces and their evolution. This original spectral imaging system is made of two cameras, a CCD (silicon Charge Coupled Device) camera for the visible spectral range (0.4-1 \textmu m) and an MCT (Mercury Cadmium Telluride) for the near-infrared range (0.8-2.5 \textmu m). The difference of resolution between the 1392 x 1040 pix VIS camera and 320 x 256 pix NIR camera mimic the typical differences of resolutions between visible colour imagers and infrared imaging spectrometers on-board planetary missions. The scene is illuminated by a monochromatic light source made of a QTH (Quartz Tungsten Halogen) lamp coupled to a multiple-gratings monochromator and a large fiber bundle pointed at the sample. The spectral resolution is controlled by the opening of two slits at the entrance and exit of the monochromator while the spectral sampling can be commanded to any arbitrary value larger than 1 nm. We refer to the spectral products as \textit{hyperspectral} when the sampling is higher than the resolution (the spectrum is fully covered with overlapping bandpasses) and as \textit{multispectral} when the sampling is lower than the resolution (discrete non-overlapping bands are selected through the spectrum).

Multi- or hyperspectral cubes are acquired by sequentially changing the bandpass of the monochromator and then taking an image with one or both camera(s). A full description of an early version of the instrument can be found in \citet{Pommerol2015106}. Compared to this earlier description, the setup has been improved to increase the spectral resolution, which is now 7 nm in the visible and 15 nm in the near-infrared. This optimisation of the Full Width at Half Maximum (FWHM) of the monochromator bandpass led to a decrease of 50$\%$ of the incident luminous flux, which we compensated for by optimising the exposure time of the cameras for each wavelength. To further enhance the signal-to-noise ratio (SNR) of the measurements, we have implemented an image averaging system. Hence, we average three pictures per wavelength in the visible and a hundred of pictures per wavelength in the near-infrared.

As noted in \citet{Pommerol2015106}, the level of the dark signal of the NIR camera is significantly influenced by room temperature. We have mitigated this impact by increasing the frequency of acquisition of dark images, which allows us to track the fluctuation of the temperature of the room. We interpolate the dark images, and we subtract the level of dark from the signal during the calibration of the data. This set of improvements has enhanced our SNR to 400 in optimal cases (i.e. bright and dry samples), as shown already in \citet{Ramy2015}.

Unfortunately, the modifications implemented to improve our spectral resolution, sampling and SNR have boosted the time needed for the acquisition of a cube, which was an average of 20 minutes in 2015 \citet{Pommerol2015106}. With the configuration used in this study, the acquisition of a cube lasts 70 minutes. To slow down the evolution of ice during this time, we cool the sample down radiatively. This is achieved by maintaining a continuous circulation of liquid nitrogen through a cylindrical shroud that surrounds the sample holder at a distance of 20 mm. 

The conditions of temperature and pressure at which each experiment was performed are specified in Table \ref{Table:mixtures}. 

\subsubsection{Light geometry}
We have performed the measurements at an incidence angle of about 20$^\circ$ and an emission angle of 30$^\circ$. The phase angle is of about 50$^\circ$ in the visible and 10$^\circ$ in the near-infrared \citep{Pommerol2015106}. We do not expect significant variations in the reflectance resulting from these different phase angles \citep{GRL:GRL53248}.

\subsubsection{Frost condensation on the samples}
The protocols developed at LOSSy guarantee frost-free intimate mixtures. Nevertheless, between the insertion of the sample holder into SCITEAS and reaching stable conditions within the chamber, the samples and sample holder can cold-trap atmospheric water. If the experiments are performed under low-pressure conditions, frost sublimates. Otherwise, it condenses again on the shroud as the samples and sample holder get warm.

\subsubsection{Ice sublimation from the samples}
The cooling system of SCITEAS is not powerful enough to stop the sublimation of surface-ice in every case, especially when working at low-pressure conditions. Sublimation decreases the ice-to-dust ratio of the first micrometres of the sample; the depths from which light scattering occurs in this spectral range. As a precaution, the concentrations of water ice in the SPIPA-B and dark basalt intimate mixtures should, therefore, be considered as upper limits.

While working on developing protocols to avoid this scenario, we have found a temporary solution in turning SCITEAS into a multispectral imager, that is, an imager with larger spectral sampling than resolution. In this configuration, we select strategic wavelengths (e.g.on the continuum and on absorption bands) at which we measure the reflectance instead of covering the whole spectral range with small steps between wavelengths. Hence, the acquisition time of each spectrum is reduced from 70 to 5 minutes. In this study, we have used this procedure with the frozen soils samples, the reflectance of which has been measured at the following wavelengths (in \textmu m) : 0.490, 0.670, 0.835, 0.985, 1.348, 1.402, 1.438, 1.498, 1.582, 1.624, 1.702, 1.852, 1.990, 1.996, 2.098 and 2.200.

\subsection{Data processing}
\label{S:3.3}
The physical calibration of the spectral data allows us to convert the raw camera units into reflectance factor (REFF) units, according to the definition of \citet{Hapke1993}. This reflectance unit is also generally referred to as I/F.cos i in the remote sensing community. According to this definition, a perfect lambertian surface would have a constant reflectance factor REFF=1 independent of the incidence, emission and phase angles. The calibration procedure consists of acquiring regularly, before and/or after the measurement of actual samples, similar measurements (same cameras settings, spectral ranges and sampling, sample position) with a large surface of Spectralon\textsuperscript{TM} (Labsphere) covering the field-of-view of the camera. This material shows a nearly Lambertian behaviour and a constant hemispheric reflectance of 0.99 over the VIS-NIR spectral range. The raw data acquired with the sample are then divided pixel-per-pixel and for every wavelength by the corresponding raw data acquired with the Spectralon surface. On average, it is enough to acquire one calibration cube per day to correct for most artefacts arising from the monochromator. 

A major issue with infrared MCT sensors in general, and our NIR camera in particular, is the high and variable thermally generated dark signal. We have mitigated this issue by installing a motorized shutter on the monochromator. Regularly during the measurements of the spectral cubes, dark images are acquired by shortly closing the shutter to acquire dark images with the exact same settings as the actual images. Temporal interpolation between the measured darks and pixel-per-pixel subtractions are then performed to correct the variations of the dark signal. This procedure is very efficient in the conditions that dark images are acquired and is usually sufficient to represent accurately the variations of the dark current.

A detailed explanation of this process is given in \citet{Pommerol2015106}. The only significant addition to the procedure is a correction for the small absorption of Spectralon\textsuperscript{TM} around 2.1-2.2 \textmu m (\citet{zhang2014}).

The calibration procedure outputs a hyperspectral reflectance cube in which we define Regions of Interest (ROIs) to extract averaged reflectance spectra from specific areas. The ROIs are defined trying to include as much of the sample as possible, without touching the pixels corresponding to the sample holders.

\subsubsection{Uncertainties}
\label{S:uncertainties}
Even though dividing the sample cube by a reference removes systematic errors, other sources of uncertainties are not accounted for. Some of these sources are the read-noise of the cameras, changes in the illumination of the sample or reference cubes due to an unexpected behaviour of the monochromator (such as partial blocking of filters or gratings), multiple reflections between the sample and the quartz window (generating ghosts and straylight) and the uncertainties related to the calibration procedure (e.g. positioning or cleanliness of the reference).

To assess the precision and accuracy of our measurements, our sample holders contain -along with the samples- a small Spectralon\textsuperscript{TM} (Labsphere) calibration target.  

To assess the precision of our measurements, we compare the reflectance spectra of each pixel within a ROI with the spatial variation of the mean spectra of that ROI (that is, we compute a standard deviation within each ROI). We average the results to obtain a relative error of 0.3$\%$ in the VIS and a 2$\%$ in the NIR, that is, signal-to-noise ratios (SNRs) of 333 and 50, respectively. 

A part of this noise is photon noise and varies as the square root of the signal that is proportional to the reflectance of the sample. Another fraction of the noise, such as read noise, is independent of the reflectance of the sample. Assuming as the worse case that the entire error found with the white reflectance target is due to photon noise, we can extrapolate to the observations of darker samples. With a reflectance of 0.1, the SNR drops to 95 in VIS (relative error of 1$\%$) and 16 in NIR (relative error of 6$\%$).

These values apply to individual pixels. When pixels are averaged within a ROI however, the photon noise decreases proportionally to the square root of the number of pixels within the ROI. 
The number of pixels that we average depends on the size of our samples. On average, our ROIs have 9000 pixels in the VIS (with a minimum of 1150 pixels and a maximum of 30000) and 800 in the NIR (with a minimum of 100 pixels and a maximum of 3000). The SNRs of our measurements increase in average to almost 10000 in the VIS and 500 in the NIR.

We assess the accuracy of our measurements by comparing the theoretical reflectance of the references with our measurements. When performing this analysis with a set of 10 measurements, we observe 6$\%$ of variability with the visible camera and 3$\%$ with the near-infrared. These values do not appear to be dependent on the signal as they show no variability with wavelength and hence the level of the signal. 

Accuracy limits, mostly, the uncertainty of our reflectance measurements. Nevertheless, the SNR and absolute accuracy have very different implications on our spectral analyses. Many spectral criteria used are relative criteria, where ratios between wavelengths of the same spectra are computed (slopes, band depths, etc., see Section \ref{S:3.3.1}). As the main limitations of the absolute accuracy (illumination homogeneity and stability) are multiplicative factors in terms of reflectance and affect all wavelengths in the same way, relative spectral criteria are independent on the absolute accuracy and are only influenced by the SNR. 

\subsubsection{Spectral analysis}
\label{S:3.3.1}
We perform a qualitative evaluation of the reflectance spectra by analysing the reflectance of the continuum, the spectral slopes of the spectra, and the shape and depth of the absorption bands. We also conduct a colour analysis based on the spectral response of the Colour and Stereo Surface Imaging System (CaSSIS) \citep{Thomas2017}. 

\paragraph{Reflectance of the continuum}
Since H$_2$O ice shows a high albedo in the visible range, the first signature to examine when looking for ice is the overall brightness of the soil. Some Mars-related studies use 1.08 \textmu m  \citep{JGRE:JGRE2291,JGRE:JGRE2906} as the wavelength of reference to study the reflectance of the soil. This choice is made for both scientific (e.g. the atmosphere shows a transmittance window) and technical reasons (e.g. the instrument used to look at the surface has a high SNR at that wavelength). In this study, we use the same wavelength for comparison purposes. Because of our 0.006 \textmu m-sampling in this range, the exact wavelength used as a reference for the continuum is 1.078 \textmu m. The maximum relative uncertainty of these measurements is 3$\%$.

\paragraph{Band depth}
Several parameters control the strength of the absorption bands \citep{JGRE:JGRE2474}. The main influences are the shape and size of the ice grains and the amount of ice present on the surface.

The definition of band depth is not unique and affects the derived values \citep{JGRE:JGRE2474}. To refer to a particular way of assessing the depth of the absorption bands and to avoid a sense of absoluteness, some authors use the term \textit{H$_2$O index} \citep{Langevin2007, Brown_2014}. This term can only be understood with the formula used to define it. For comparison purposes, we work with formulae already in use by other authors.

To assess the water index around 1.5 \textmu m, we use the formula proposed by \citet{JGRE:JGRE2291}, often employed in Martian spectrometry studies \citep{JGRE:JGRE2906, JGRE:JGRE2922,Brown_2014} and reproduced in Equation \ref{water_depth}.

\begin{equation}
\label{water_depth}
Water\:index(1.5 \: \mu m) = 1 - \frac{R(\lambda _0)}{R(\lambda _1)^{0.7} \times R(\lambda _2)^{0.3}}
\end{equation}

In this formula, the shoulders of the bands are fixed at $\lambda _1$  = 1.385 and $\lambda _2$  = 1.772 \textmu m, since those wavelengths present weak atmospheric CO$_2$ signatures. Using the sampling of SCITEAS, we have set the shoulders at $\lambda _1$  = 1.384 and $\lambda _2$ = 1.774 \textmu m. For the 2.0 \textmu m-index, we work with the definition proposed by \citet{masse2008mineralogical} and reproduced in Equation \ref{water_depth2}. 

\begin{equation}
\label{water_depth2}
Water\:index(2.0 \: \mu m) = 1- \frac{R(\lambda _0)}{\frac{W_1}{W} \times R(\lambda _2) + \frac{W_2}{W} R(\lambda _1)}
\end{equation}

Where $\lambda _1$ = 1.834, $\lambda _2$ = 2.248 \textmu m, W=($\lambda _2$ - $\lambda _1$), W$_1$=($\lambda _0$ - $\lambda _1$) and W$_2$=($\lambda _2$ - $\lambda _0$). For both formulae, $\lambda _0$ refers to the position of the reflectance minimum within the band (that is, not at 1.5 and 2.0 \textmu m systematically).

To assess whether a band is present, we trace the continuum (as a line) between the previously mentioned shoulders: if the majority of the reflectance points between the shoulders fall under that linear fit, then we assume the existence of a band. 

\begin{figure}
\centering
\includegraphics[width=\textwidth]{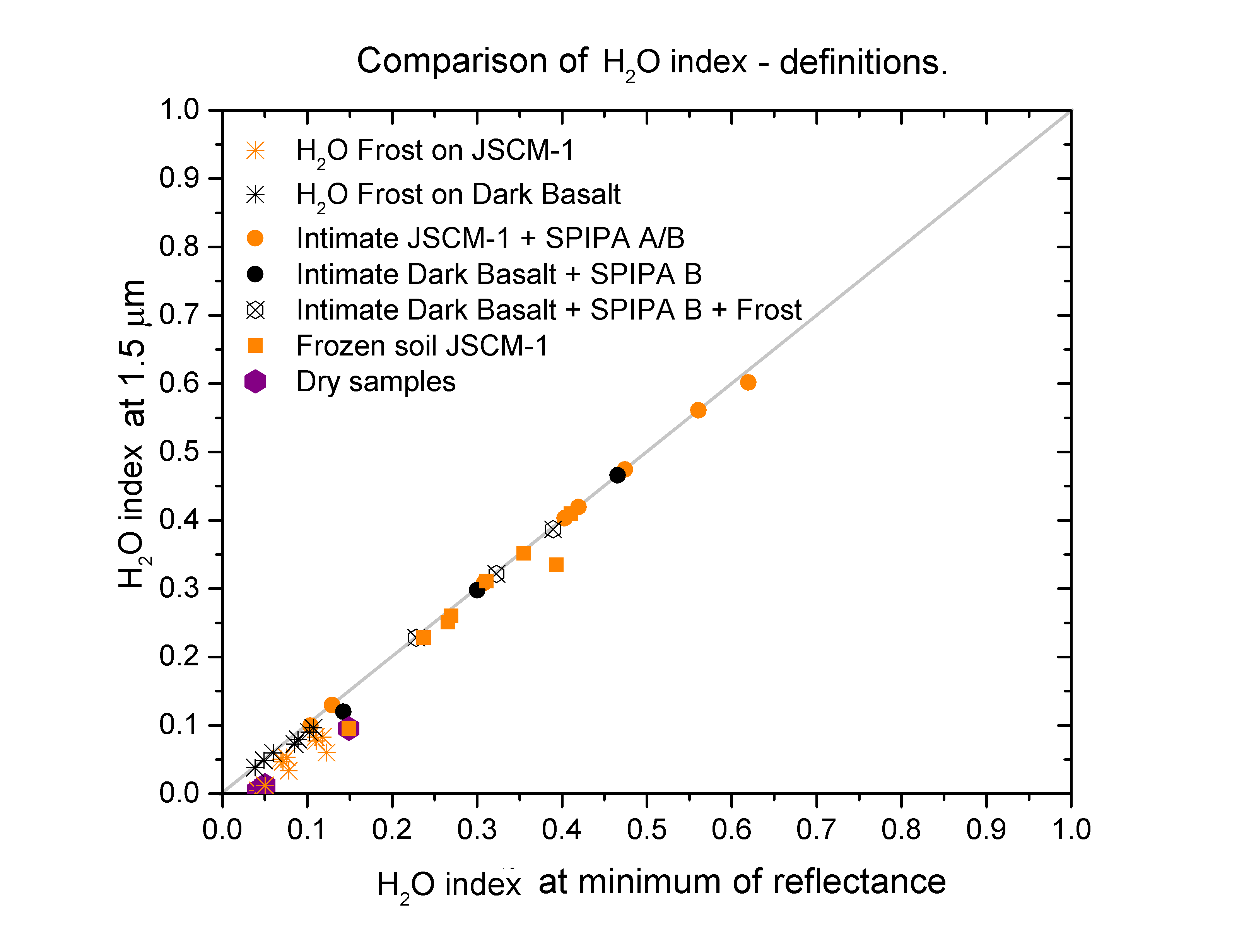}
\caption{(Comparison of the water indexes obtained by assessing them at 1.5 \textmu m (y-axis) and the minimum of reflectance of the band (x-band). The grey line indicates the perfect correlation.}
\label{Fig:comparison_bands}
\end{figure}

The shifts of the positions of the minima of the bands introduce errors in the evaluation of the water indices when assessed at 1.5 and 2.0 \textmu m instead of at the minimum of reflectance in hydrated minerals. Figure \ref{Fig:comparison_bands} shows that this only happens when there is not enough water ice to shift the band towards 1.5 and 2.0 \textmu m, that is, in the experiments with frost. From Figure \ref{Fig:comparison_bands} we deduce that, in the worst case scenario, measuring the water index at 1.5 \textmu m can lead to underestimates of their values by up to 50$\%$. 

The uncertainty in the derived depth of a band propagates from the uncertainties on the reflectance values used in its calculation. As the band depth is defined as a ratio of two values of reflectance, the relative uncertainty of the band depth is equal to the sum of the relative uncertainties of the two values of reflectance used in the calculation. The absolute accuracy, that affects the entire spectra with a multiplicative factor and independent on the wavelength, does not affect the band depth. The uncertainty on the band depth is therefore only affected by the SNR.

\paragraph{Position of barycentre of the bands}
Adding hydrated minerals to pure water ice introduces an asymmetry into the absorption band of water at 2.0 \textmu m \citep{JGRE:JGRE2559}, and a shift in the positions of the minima of both bands at 1.5 \textmu m and 2.0 \textmu m. We have evaluated the shift of the shape of the bands by tracking their barycentres (a.k.a. centroids). 

To define the barycentre, we first define the continuum as a straight line between the shoulders of a band. Then, we apply Equation \ref{centroid}, where {\textlambda} stands for the wavelength, R$_C$(\textlambda) is the reflectance of the continuum at that wavelength and R(\textlambda) is the reflectance measured at that wavelength. We have used the same shoulders as for the calculation of the band depths: 1.384 and 1.774 \textmu m for the absorption band at 1.5 \textmu m, and 1.834 and 2.248 \textmu m for the band at 2.0 \textmu m. 
 
\begin{equation}
\label{centroid}
Barycentre = \frac{\sum_{i=0}^n \lambda _i (R_C(\lambda_i) - R(\lambda _i))}{\sum_{i=0}^n (R_C(\lambda_i) - R(\lambda _i))}
\end{equation}

\paragraph{Spectral Slope}
We have determined the relative spectral slope of our samples both in the visible and in the near-infrared following Equation \ref{slope}. 
\begin{equation}
\label{slope}
S_r = \frac{R_{\lambda _2}- R_{\lambda _1}}{R_{\lambda _1} (\lambda _2 - \lambda _1)} \times 10^4 \quad \quad (\% /100 nm)
\end{equation}
$\lambda _1$ and $\lambda _2$ have been set to 0.445 and 0.745 \textmu m in the visible and 1.1 and 1.7 \textmu m in the near-infrared. R$_{\lambda _1}$ and R$_{\lambda _2}$ refer to the values of reflectance at $\lambda _1$ and $\lambda _2$ respectively.

As for the band depth calculation, the relative uncertainties are therefore only affected by the SNR. 

\subsubsection{Colour analysis}
\label{S:3.3.2}

\begin{figure}
  \centering
  \includegraphics[width=\textwidth]{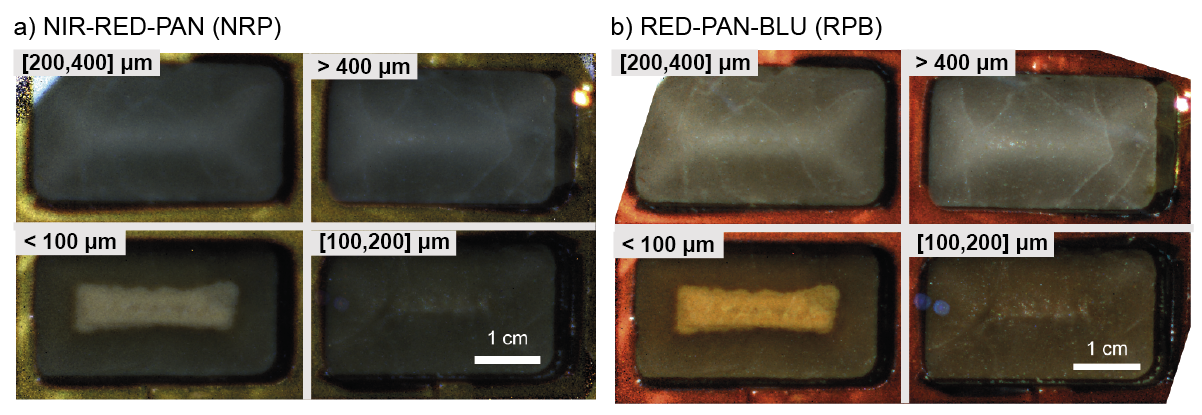}
  \caption{CaSSIS colour composites of Figure \ref{Fig:mixtures}d, that is, JSC Mars-1 frozen soils. Labels refer to the size fraction of JSC Mars-1. a) NIR-RED-PAN composite b) RED-PAN-BLU composite.}
  \label{Fig:cassis}
\end{figure}

CaSSIS is the high signal-to-noise ratio (SNR) camera on board ESA's Trace Gas Orbiter (TGO). TGO follows a non-Sun-synchronous orbit around Mars. This type of orbit allows us to observe diurnal phenomena, and therefore is of particular interest for frost detection. For this reason, we use as colours of reference CaSSIS' four bands BLU, PAN, RED and NIR (which have their nominal effective centre at 485, 675, 840 and 985 nm respectively \citep{Thomas2017}).

We have convolved the signal measured from our samples with the spectral response of CaSSIS to obtain a colour image of our sample as it would have been seen by CaSSIS. All technical and calibration details of CaSSIS needed to convolve our signal have been published in \citet{Roloff2017}. Below, we explain the steps taken to simulate what we have called \textit{the CaSSIS colours}.

\begin{enumerate}
\item We first scale the solar radiance to an average Sun-Mars distance of 1.52 a.u. Then, we multiply it by the cosine of the simulated incidence angle (we have chosen i=60$^\circ$), the angular aperture of the telescope (i.e, 0.023 rad) and by the area of the primary mirror, which has a radius of 0.07 meters.
\item We divide the total power received by the sensor by the number of pixels and multiply it by the exposure time.  
\item The previous result is divided by the throughput of the telescope, whose spectral characteristics are determined by the reflectivity of the mirror. The number of photons received is converted to electrons through the quantum efficiency.
\item For each colour filter, we calculate the instrument spectral response and the electrons we would detect for a lambertian white surface (maximum amount of electrons).
\item We multiply the VIS-NIR reflectance measured with SCITEAS with the spectral response of each filter and normalise it by the maximum amount of electrons (analogue to Lambertian surface) to have units of REFF. We can now produce colour images such as the ones in Figure \ref{Fig:cassis}, which shows NIR-RED-PAN and RED-PAN-BLU CaSSIS composites of the JSC Mars-1 frozen soils shown in Figure \ref{Fig:mixtures}d.
\item We extract the reflectance in a filter of a given area by defining ROIs, as performed for the spectral analysis. The different BLU/PAN/RED/NIR signals can be combined to retrieve different band ratios. 
\end{enumerate}

In order to assess the uncertainties of the CaSSIS colours, we have proceeded in the same way as in Section \ref{S:uncertainties}, that is, we have analysed the reflectance of the reference sample (Spectralon) in the BLU, PAN, RED and NIR products. On average, we have obtained an uncertainty in the precision and accuracy of 4.5$\%$ and 0.58$\%$ respectively. With a mean of 9000 pixels averaged per ROI, the SNR of the derived colours is over 500, for both bright and dark samples. 

\subsection{Samples}
\label{S:3.1}
\begin{figure}
  \centering
  \includegraphics[width=1.25\textwidth, angle=90]{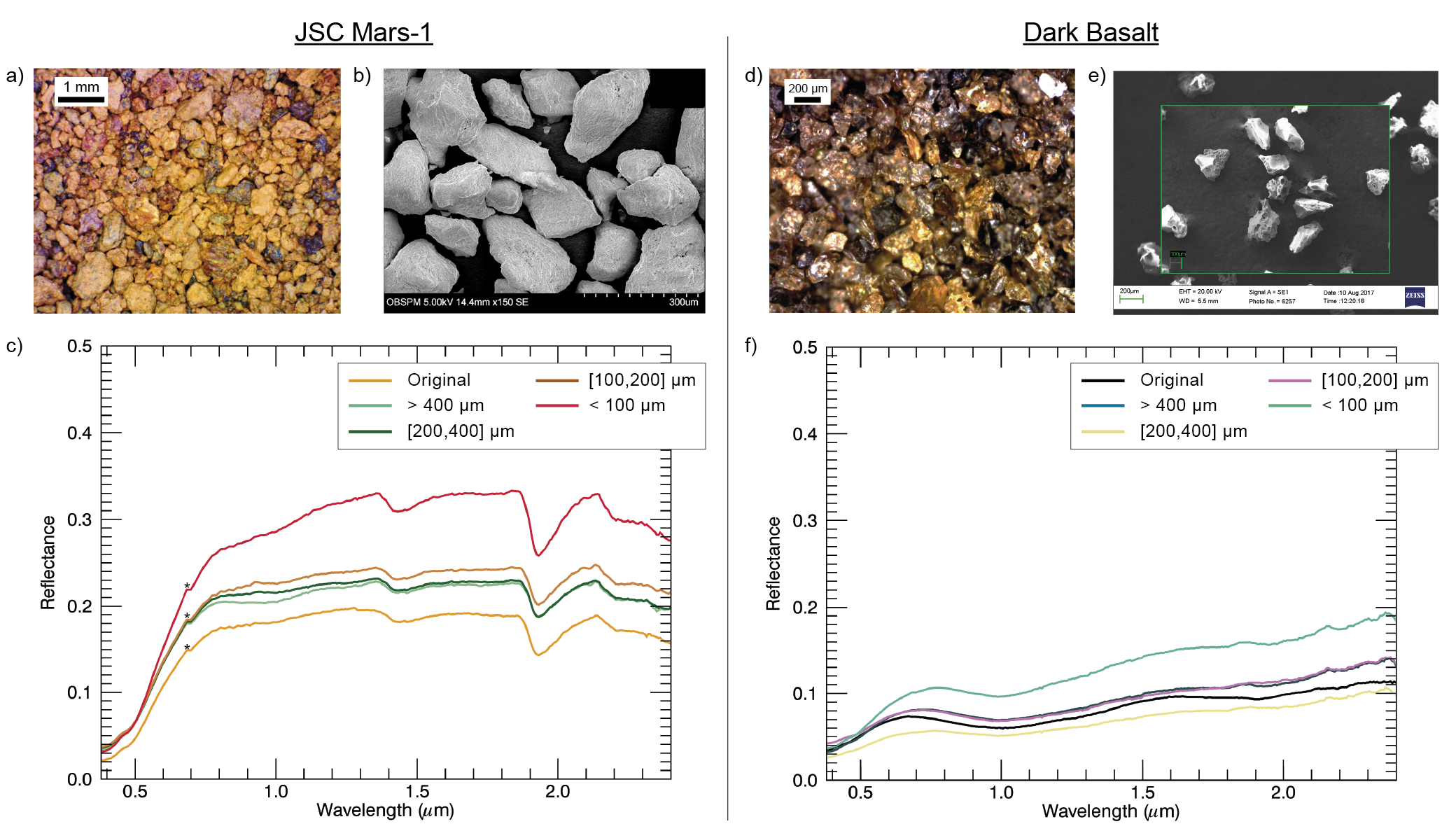}
  \caption{Soil simulants used in the experiments of this study. a) Optical microscope image of JSC Mars-1 b) SEM image of JSC Mars-1 (80-200 \textmu m fraction) reproduced from \citet{Yann:thesis} c) Reflectance spectra of the size distributions of JSC Mars-1 d) Optic microscope image of dark basalt e) SEM image of dark basalt, acquired in the Institute of Geography of the University of Bern f) Reflectance spectra of the size distributions of the dark basalt. Stars indicate artefacts from the calibration process.}
  \label{Fig:samples_dry}
\end{figure}

\subsubsection{Martian analogues}
\label{S:3.1.1}
JSC Mars-1 is the Martian regolith simulant distributed by NASA Johnson Space Center (see Figure \ref{Fig:samples_dry}, left). As explained in \citet{Allen:1997}, JSC Mars-1 is the $<$1 mm fraction of weathered volcanic ash from Pu’u Nene, a cinder cone on the Island of Hawaii. Even though JSC Mars-1 matches the visible and near-infrared reflectance spectra of the bright regions of Mars, it is more hydrated than actual Martian soil (see Figure 1 in \citet{Allen:1997}). In terms of particle size, JSC Mars-1 ranges from 1000 \textmu m down to less than 20 \textmu m \citep{Allen:1997}. To vary the size of grain of JSC Mars-1, we dry sieved it to obtain four distribution sizes: particles bigger than 400 \textmu m; particles between 200 \textmu m and 400 \textmu m; particles between 100 \textmu m and 200 \textmu m, and particles smaller than 100 \textmu m. From now on, we will refer to these size fractions as $>$400 \textmu m; [200,400] \textmu m; [100,200] \textmu m and $<$ 100 \textmu m respectively. Figures \ref{Fig:samples_dry}a, \ref{Fig:samples_dry}c and \ref{Fig:samples_dry}d show optical and scanning electron microscope images of non-sieved JSC Mars-1 ---extracted from \citet{Yann:thesis}---, as well as the reflectance spectra of the different size distributions. Only the smallest size distribution shows a noticeable rise in reflectance; up to 50$\%$ in the near-infrared. The reflectance of the original size distribution is lower than the one of the sieved fractions, which may be due to different preparations of the samples, as already studied by \citet{JGRE:JGRE20158}. The reflectance we measured agrees with the one published therein. 

Other regions of Mars show lower albedo along with a basaltic composition \citep{Bandfield:2002, Christensen:2003, Bibring:2005, Poulet:2007}. We have therefore used, as a second analogue, a dark, basaltic sample. This sample was collected from active dunes in the Ka’u Desert region in Hawaii by C. Okubo and will be further on referred to as \textit{dark basalt}. More information about this material can be found in \citet{JGRE:JGRE20158}. 

We sieved the sample to obtain the same size distributions used with the JSC Mars-1, that is, $>$400 \textmu m; [200,400] \textmu m; [100,200] \textmu m and $<$ 100 \textmu m. Figures \ref{Fig:samples_dry}d, \ref{Fig:samples_dry}e and \ref{Fig:samples_dry}f show optical and scanning electron microscope images of the non-sieved dark basalt, together with the reflectance spectra of the different size distributions.

\subsubsection{Ice}
\label{S:3.1.2}

\begin{figure}
  \centering
  \includegraphics[width=\textwidth]{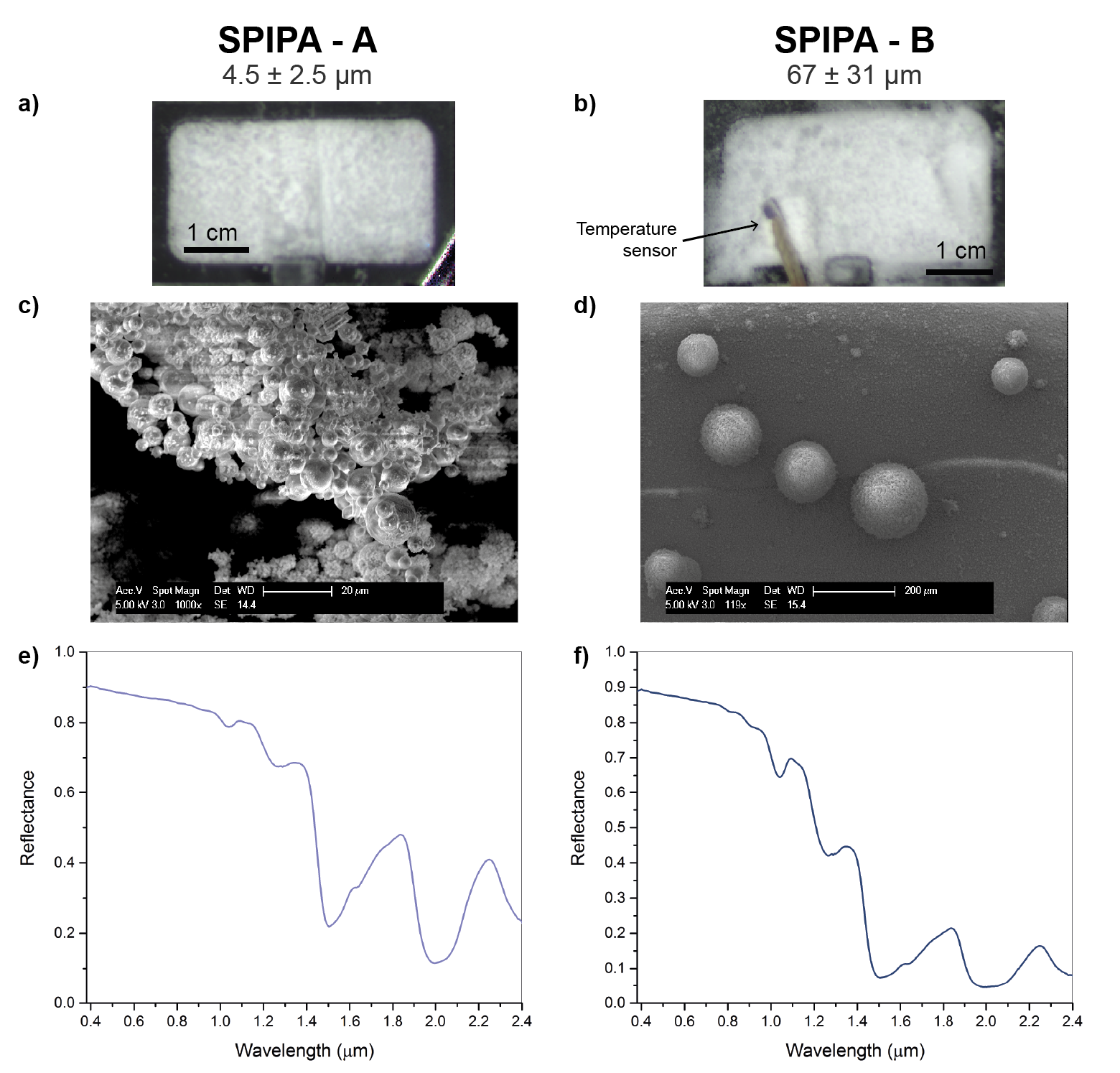}
  \caption{Water ice used in the intimate mixtures. a) and b) Pictures of SPIPA-A (left) and -B (right) in a sample holder c) and d) Cryo-SEM images of SPIPA-A (left) and -B (right) acquired at the CSEM (Switzerland) e) and f) Reflectance spectra of SPIPA-A (left) and -B (right) }
  \label{Fig:samples_ice}
\end{figure}

We have used the Setups for Production of Icy Planetary Analogues (SPIPA) at LOSSy \citep{Pommerol:2019} to produce the particulate water ice used in the preparation of intimate mixtures. In this study, we have used two grain distributions, which will be referred to as SPIPA-A and SPIPA-B. SPIPA-A are spherical water ice particles with a mean diameter of 4.5 \textmu m and a standard deviation of 2.5 \textmu m, while SPIPA-B consists of particles that have a mean diameter of 67 \textmu m with 31 \textmu m of standard deviation. Both SPIPA ices were produced by freezing nebulised, deionised water. We obtained different size distributions by using two different systems to nebulise the water. For details on the preparation of the SPIPA ices, see \citet{Pommerol:2019}. The diameters of the ice particles have been measured from cryo-scanning electron microscope (cryo-SEM) images. Surface water ice on Mars is expected to be in the micron range \citep{Cull:2010}, except for the ice grains in the polar caps, which can be millimeter sized \citep{Barr_2008}. Figures \ref{Fig:samples_ice}a and \ref{Fig:samples_ice}b show pictures of sample holders filled with SPIPA A and B. Figures \ref{Fig:samples_ice}c and \ref{Fig:samples_ice}d show cryo-SEM pictures of both type of ices. Finally, we show in Figures \ref{Fig:samples_ice}e and \ref{Fig:samples_ice}f the reflectance spectra of the pure samples. 

\subsubsection{Ice-dust mixtures.}
\label{S:3.1.3}

\begin{figure}
  \centering
  \includegraphics[width=\textwidth]{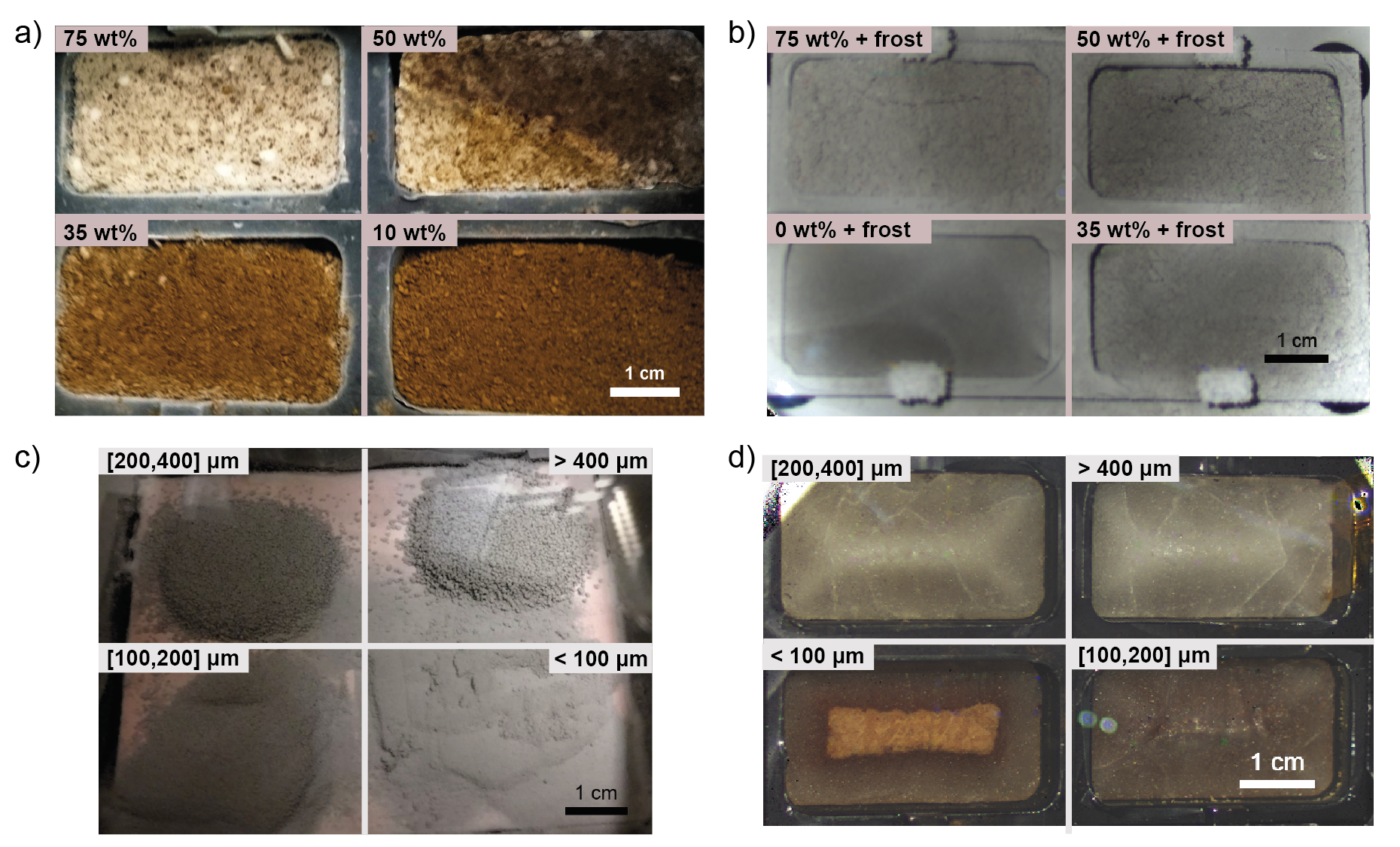}
  \caption{Optical images of some of the mixtures characterised in this study. a) JSC Mars-1 and SPIPA-A intimate mixtures. The weight percentage of water ice is indicated next to each sample. The picture with 50 wt$\%$ ice shows a shadow; the mixture is homogeneous  b) Frost over intimate mixtures of Dark Basalt and SPIPA-B ice. The weight percentage of water ice contained in the intimate mixtures is indicated next to each sample c) Step 3 of frost deposition on different size distributions (indicated next to each sample) of dark basalt d) Frozen soils of different size distributions (indicated next to each sample) of JSC Mars-1 after 17h of sublimation. The ice of the bottom left sample holder has sublimated, revealing the dust underneath, which appears on the picture as a bright rectangle.}
  \label{Fig:mixtures}
\end{figure}

We have produced three different types ice-dust mixtures: frost, intimate mixtures, and frozen soils. A view of the samples can be found in Figure \ref{Fig:mixtures}. 
{\renewcommand{\tabularxcolumn}[1]{m{#1}}
\newcolumntype{s}{>{\hsize=.5\hsize}X}
\setlength\tabcolsep{3pt}
\begin{landscape}
\begin{table}[!p]
\tiny
\begin{tabularx}{\hsize}{s s s s s s X s s}
\hline
\textbf{Mixing \newline Mode} & \textbf{Water ice}& \textbf{Regolith Simulant} &   \textbf{Size of \newline  regolith} &  \textbf{Ice \newline Content} & \textbf{Acquisition \newline mode} & \textbf{Time of acquisition} & \textbf{T$_{Shroud}$ \newline (K)} & \textbf{P \newline (mbar)}\\
\hline
\multirow{8}{*}{\shortstack[l]{Frost \\ Deposition}} & \multirow{8}{1.5cm}{Condensed atmospheric \newline water} & \multirow{4}{*}{JSC Mars-1 } & $>$400 \textmu m & \multirow{8}{1.5cm}{See \newline Table \ref{Table:Frost_thickness}}  &  \multirow{8}{1.5cm}{Hyperspectral \newline 0.4-2.4 \textmu m} & \multirow{8}{*}{ \shortstack[l]{Step 1: 10 min condensation \\+ 70 min measurement \\ Step 2: Step 1 \\+ 10 min cond.\\ + 70 min measurement \\ Step 3: Step 2 \\+ 10 min cond. \\+ 70 min meas.}} & \multirow{4}{*}{$\sim$110} & \multirow{8}{*}{1000}\\ 
& & & [200,400] \textmu m  &   &  \\
& & & [100,200] \textmu m  &  & \\
& & & $<$100 \textmu m  &  & \\
\cline{3-4}
\cline{8-8}
& &\multirow{4}{1cm}{Dark \newline Basalt } & $>$400 \textmu m & & & & \multirow{4}{*}{$\sim$100}\\ 
& & & [200,400] \textmu m  &   &  \\
& & & [100,200] \textmu m  &  & \\
& & & $<$100 \textmu m  &  & \\
\hline
\multirow{3}{*}[-10pt]{\shortstack[l]{Intimate \\ mixture}} & SPIPA-A & \multirow{1}{*}{JSC Mars-1} & \multirow{2}{*}[-5pt]{\shortstack[l]{71.7$\%>$ 150\textmu m \\ 28.3$\%<$ 150\textmu m$^{1}$}} & 10wt$\%$,35wt$\%$ 50wt$\%$,75wt$\%$ &\multirow{3}{*}[-10pt]{\shortstack[l]{Hyperspectral\\ 0.4-2.4 \textmu m}} & \multirow{3}{*}[-10pt]{70 min} & $\sim$90 & $\sim$20\\ 
\cline{2-3}
\cline{5-5}
\cline{8-9}
& \multirow{2}{*}{SPIPA-B} & JSC Mars-1 & &10wt$\%$,35wt$\%$ 50wt$\%$ 75wt$\%$ & &  & $\sim$100 & $\sim$20\\
\cline{3-5}
\cline{8-9}
& & Dark Basalt & 50$\% >$ 109\textmu m \newline 50$\% <$ 109\textmu m  & 35wt$\%$,50wt$\%$ 75wt$\%$ & & & $\sim$120 & $\sim$1\\
\hline
{\shortstack[l]{Intimate \\ mixture \\ + Frost}} & {\shortstack[l]{SPIPA-B + \\ Atmospheric \\water}} & {\shortstack[l]{Dark \\ Basalt }} & {\shortstack[l]{50$\% >$ 109\textmu m \\ 50$\% <$ 109\textmu m }} & {\shortstack[l]{35wt$\%$ + Frost\\ 50wt$\%$ + Frost\\ 75wt$\%$ + Frost}} & {\shortstack[l]{Hyperspectral \\ 0.4-2.4 \textmu m}} & {70 min} & {$\sim$120} & {1000}\\ 
\hline
\multirow{9}{*}{ \shortstack[l]{Water \\ saturated \\ Frozen \\ soil}} & \multirow{10}{1.5cm}{Slab} & \multirow{4}{*}{JSC Mars-1 } &
\multirow{4}{*}{$<$100 \textmu m } & \multirow{4}{*}{87wt$\%$} &  \multirow{4}{2cm}{Int.1: x3 Hyper \newline Int.2: x2 Hyper \newline Int.3: x4 Hyper \newline Int.4: x1 Hyper} & \multirow{4}{1.5cm}{Int.1: 0-3h \newline Int.2: 4-7h \newline Int.3: 9-14h \newline Int.4: 17h} & \multirow{4}{*}{\shortstack[l]{ Int.1: 90-140 \\ Int.2: 160 \\ Int.3: 220-260 \\ Int.4: 280}} & \multirow{4}{*}{\shortstack[l]{ Int.1:70-2e$^{-4}$ \\ Int.2: 2e$^{-4}$ \\ Int.3:0.01-0.03 \\ Int.4: 2}}\\ 
& & & &  &  \\
& & & & & \\
& & & &  & \\
\cline{3-9}
& &\multirow{5}{1cm}{Dark \newline Basalt } & \multirow{5}{1.5cm}{$<$100 \textmu m} & \multirow{5}{1.5cm}{90wt$\%$} & \multirow{5}{2cm}{Int.1: x2 Hyper \newline Int.2: x4 Multi \newline Int.3: x4 Multi \newline Int.4: x4 Multi \newline Int.5: x2 Multi} & \multirow{5}{2cm}{Int.1: 0-3h \newline Int.2: 3-3.5h \newline Int.3: 3.5-4h \newline Int.4: 4-4.5h \newline Int.5: 4.5-4.6h}& \multirow{5}{2cm}{Int.1: $\sim$125 \newline Int.2: 150-190 \newline Int.3: $\sim$195 \newline Int.4: 195-210 \newline Int.5: 210-235} & \multirow{5}{2cm}{Int.1: 70-4e$^{-4}$ \newline Int.2: 5e$^{-4}$ \newline Int.3: 6e$^{-4}$ \newline Int.4: 6e$^{-4}$-0.01 \newline Int.5: 0.015}\\
& & &  &   &  \\
& & &   &  & \\
& & &   &  & \\ 
& & & & &  \\
\hline
\end{tabularx}
\caption{Summary of the composition, concentrations, mixture types and acquisition conditions of the experiments reported in this study. The abbreviation \textit{Int.} stands for \textit{interval}.  $^{1}$\citep{Allen:1997}}
\label{Table:mixtures}
\end{table}
\end{landscape}}

\paragraph{Intimate mixtures}
In intimate mixtures, the grains of the end-members are mixed to the particle scale \citep{Poch2016154}. At LOSSy, we work with protocols designed to obtain homogeneous and reproducible mixtures that have already been shown to be successful in studies such as \citet{GRL:GRL53248} and \citet{Poch2016154,Poch2016288}. These papers also provide detailed information about the mixing procedure. In summary, both materials are blended with the help of a vortex mixer while they are repeatedly plunged into liquid nitrogen to avoid sublimation and metamorphism of ice during the process. 

Because porosity influences the reflectance of the samples, we always fill the sample holders without compressing the samples: the sample holders are first overfilled and then the excess is removed with a spatula. In this way, we obtain non-compacted samples with flat surfaces. For reference, the porosity of non-compacted SPIPA-A ice is 80.6 $\pm$ 0.9 $\%$ and the one of non-compacted SPIPA-B, 49.3 $\pm$ 1.4 $\%$ \citep{Brouet2016}.

We have found difficulties, however, in mixing the dark basalt with SPIPA-A; we could not combine the dark basalt homogeneously with such small ice particles because the basalt tended to group in conglomerates. We also noticed rapid warming of the dark basalt grains, probably caused by their low albedo, which provoked the sublimation of the surface ice before the entire acquisition of the hyperspectral cube. This problem appeared as well when producing intimate mixtures of dark basalt with only 10 wt$\%$ of SPIPA-B. For that reason, neither intimate mixtures of dark basalt with SPIPA-A nor with a 10 wt$\%$ of SPIPA-B are shown in this study.

Figure \ref{Fig:mixtures}a shows four sample holders filled up with intimate mixtures of JSC Mars-1 and SPIPA-A. The sample holders are rectangular containers of dimensions 2x4x1 cm. The different contents of ice, ranging from 10 wt$\%$ to 75 wt$\%$, are indicated next to each sample holder. An overview of the ice-to-dust ratios of the mixtures analysed in this paper is given in Table \ref{Table:mixtures}.

\begin{figure}
  \centering
  \includegraphics[width=\textwidth]{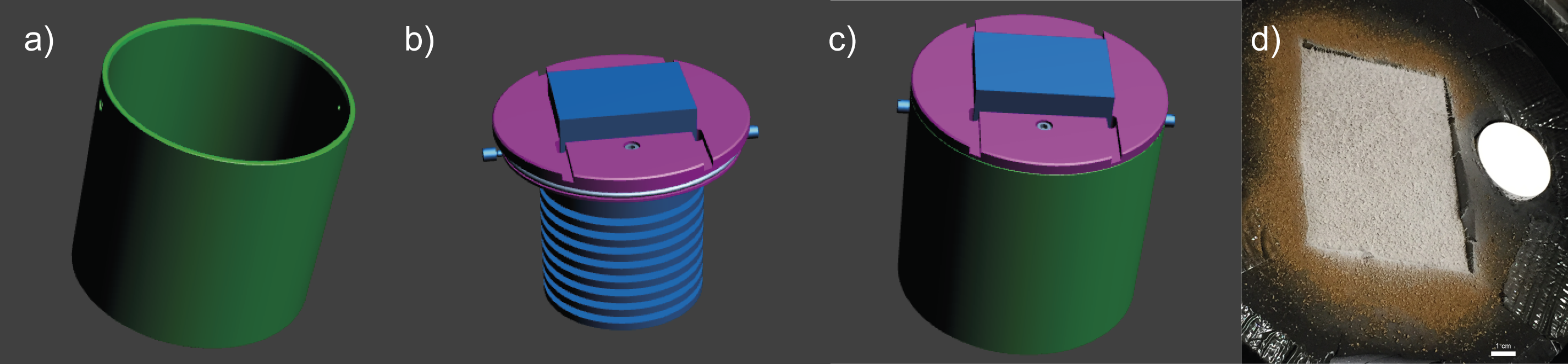}
  \caption{Sample holder to cold-trap frost. The sample holder consists of two pieces; a) a cylinder that contains liquid nitrogen and b) the cover of the cylinder pierced by a copper piece. c) Put together, the copper disks are plunged into liquid nitrogen. The powder is spread in the blue part in c), which remains cold enough for atmospheric water to condense on it. d) Example of frost condensing over the JSC Mars-1 powder in contact with the copper surface. An insulating black foam covers the rest of the remaining metallic surface.}
  \label{Fig:frost_trap}
\end{figure}

\paragraph{Water Frost}
Throughout this study, we refer to frost as the low-density, fresh ice that results from both the direct condensation of atmospheric water onto a cold surface and the deposition onto a surface of ice particles condensed in the atmosphere above the surface. To simulate the appearance of frost onto a granular surface, we have designed and constructed a sample holder that allows cold-trapping of atmospheric water onto a specific surface. 

The sample holder consists of two parts (Figure \ref{Fig:frost_trap}). The first part is a cylindrical container made of aluminium (Figure \ref{Fig:frost_trap}a). The second part is the container's cover, coloured in pink in Figure \ref{Fig:frost_trap}b. This cover is perforated by a rectangular, copper-made piece, coloured in blue in Figure \ref{Fig:frost_trap}b. Above the cover, the copper piece is a cuboid of 8x6x2 cm. Below the cover, the cuboid vertebrates an array of discs.  To create a cold trap, we fill up the cylinder (Figure \ref{Fig:frost_trap}a) with liquid nitrogen and cover it with the element shown in Figure \ref{Fig:frost_trap}b, bringing the copper element (coloured blue in Figures \ref{Fig:frost_trap}b and \ref{Fig:frost_trap}c) in contact with the liquid nitrogen. As a result of the high thermal conductivity of copper and the large contact surface of the copper discs with the liquid nitrogen, the entire copper element is kept at very cold temperatures. Hence, the exposed part of the copper element remains below 220 K for an hour when in contact with an atmosphere at around 290 K. Figure \ref{Fig:frost_trap}c shows the sample holder once its two parts are mounted together. We then spread the dusty sample over the copper surface in layers of about 2 mm to ensure good thermal contact of the dust and the surface, minimising thermal gradients inside the layer. We cover the rest of the sample holder with an isolating foam to make sure that water condenses preferentially on the sample in contact with the metallic surface (black area in Figure \ref{Fig:frost_trap}d). 3D model files of this sample holder are available to share under request.

When we insert the sample holder into our simulation chamber, we use the lid of the chamber to control the deposition of frost; we open the lid to expose the sample to the atmosphere and close it to limit the water vapour available to condense. Because of the constant sublimation of liquid nitrogen from inside the sample holder, the atmosphere around the surface remains totally dry when the lid is closed and very dry when the lid is open as the sublimated nitrogen expands into the atmosphere of the lab, leaving only minute amounts of water vapor to reach the cold surface. This limits the frost growth rates to very low values and mimics the Martian situation with a very cold surface and very dry atmosphere, although the major inert gas is nitrogen in the lab and CO$_2$ on Mars. Away from the simulation chamber, the atmosphere of the lab has an average relative humidity of $\sim$30$\%$ and temperature of 293 K.

In this article, we show several steps of frost deposition on the samples; each step equates to ten minutes of exposure to the atmosphere followed by the acquisition of a hyperspectral cube. Figure \ref{Fig:mixtures}c shows the state of the different size distributions of dark basalt after the third step of deposition.

The amount of frost that condenses on a sample at a given pressure depends on the conditions of humidity in the room and the temperature of the sample, which in turn depends on the properties of the material (heat capacity and grain-size). Thus, when we expose different size fractions of the same material to the atmosphere, not every size fraction necessarily cold-traps the same amount of water. Therefore, in this study, it is more straightforward to compare spectra of different steps of frost deposition on a single size distribution than it is to compare the same deposition-steps between different size distributions.

With the current setup of SCITEAS, we do not have the means of quantifying the amount and thickness of frost condensed onto the samples, especially during early stages of deposition. The 10 minutes of exposure were chosen because this duration produces a small but recognizable change of albedo by eye. The results of \citet{Clark:1981} however, offer us a possibility to estimate indirectly the thickness of frost deposited. Indeed, \citet{Clark:1981} measured both the frost thickness and the near-infrared spectra of two samples for several steps of frost deposition. The samples, red cinder from Mauna Kea and charcoal have high contrast in albedo. The analysis of these data show that for the first two to three depositions of frost (first 100 to 150 \textmu m of frost), the band depth at 2 \textmu m shows a nearly linear relation with frost thickness, independent of the nature and albedo of the sample. As our two analogues fall in between the samples studied by \citet{Clark:1981} in terms of albedo and colour, we assume that this relation is also true for our samples and apply it to estimate our deposited frost thickness (Thickness [\textmu m] = 3.2 + 412.9 x BD(2 \textmu m), where BD(2 \textmu m) stands for the band depth at 2 \textmu m). Note that as our thickness estimates rely entirely on the results of \citet{Clark:1981}, they should not be considered as a confirmation of the previous work or as a new way of relating the spectro-photometric data to the frost thickness. For future experiments, an independent way of measuring accurately the thickness of the frost to verify this relation, as well as other properties of the frost layer such as density, texture... would be highly desirable.

Interestingly, no particular trend between deposited frost thickness and particle size of the substrate was observed for our two samples. To reduce the dispersion in values, we therefore averaged all four size fractions to obtain the averaged estimates of deposited frost thickness shown in Table \ref{Table:Frost_thickness}. Given the uncertainties and potential differences in the procedures used by \citet{Clark:1981} and our own procedures, these values should be considered as indicative estimates rather than precise data. 

 \begin{table}[t]
     \centering
     \begin{tabular}{c | c | c | c}
         \hline
         \textbf{Sample} & \textbf{First Step} & \textbf{Second Step} & \textbf{Third Step}\\
         \hline
          JSC Mars-1 & 14 $\mu$m & 33 $\mu$m & 50 $\mu$m \\
          Basalt & 13 $\mu$m & 52 $\mu$m & 75 $\mu$m \\
          Basalt + SPIPA-B & 80 $\mu$m & -- & -- \\
         \hline
     \end{tabular}
     \caption{Averaged estimates of deposited frost thickness onto different samples.} 
     \label{Table:Frost_thickness}
 \end{table}

We also present in this study reflectance measurements of intimate mixtures of dark basalt and SPIPA-B covered by water frost (Figure \ref{Fig:mixtures}b). This sample was produced unwittingly when, at the time we were about to measure the reflectance of the intimate mixtures, our vacuum pump stopped working correctly. Consequently, the water of the atmosphere in the chamber condensed on the sample. We acquired the reflectance cube of the intimate mixtures with frost before resuming the pumping and removing this superficial frost by sublimation. This serendipitous measurement is shown to be rather useful. From the work \citet{Clark:1981}, we estimate the amount of frost deposited onto our samples to be around 80 $\mu$m.

\paragraph{Frozen soil}
We prepared what we refer to as \textit{frozen soils} by pouring deionised water onto the powders until the pores saturated and a layer of water appeared on top of the dusts. The sample holders used for these samples were slightly thicker than the ones previously used, i.e. 2x4x2 cm. For each dust, we produced four different samples, one per size distribution. The sample holders were then introduced into a freezer (230 K) and left to freeze through the night. Because of the expansion of the ice, the samples lost the flat surfaces they used to have when water was in its liquid state. The frozen soils were exposed to low pressures to monitor the evolution of their reflectance as ice sublimated. Hence, we observed the sample change from the form of \textit{ice over soil} to one in the form of \textit{ice in soil}. The conditions of pressure and temperature at which the spectra were acquired are specified in Table \ref{Table:mixtures}.

Figure \ref{Fig:mixtures}d shows an example of JSC Mars-1 frozen soils after 17h of sublimation. It can be noticed that the ice on top of the finest fraction (bottom-left sample holder, with dust smaller than 100 micrometres) has already sublimated, revealing the regolith inside. The other sample holders still have a layer of ice covering the regolith. There were no spectral differences in the evolution of the reflectance of the samples with dusts of different sizes. Therefore, we only show here the spectra of the sample with the finest dust, where the dust was revealed by the end of the experiment. In some of the sample holders, e.g. the top right one, cracks are still visible. Since particles with lower density floated in the water, the ice on top of the samples was not pure.

\section{Results}
\label{S:4}
\subsection{Reflectance spectra}
\label{S:4.1}
\subsubsection{Frost}
\label{S:4.1.1}
\begin{figure}
  \centering
  \includegraphics[width=0.84\textwidth]{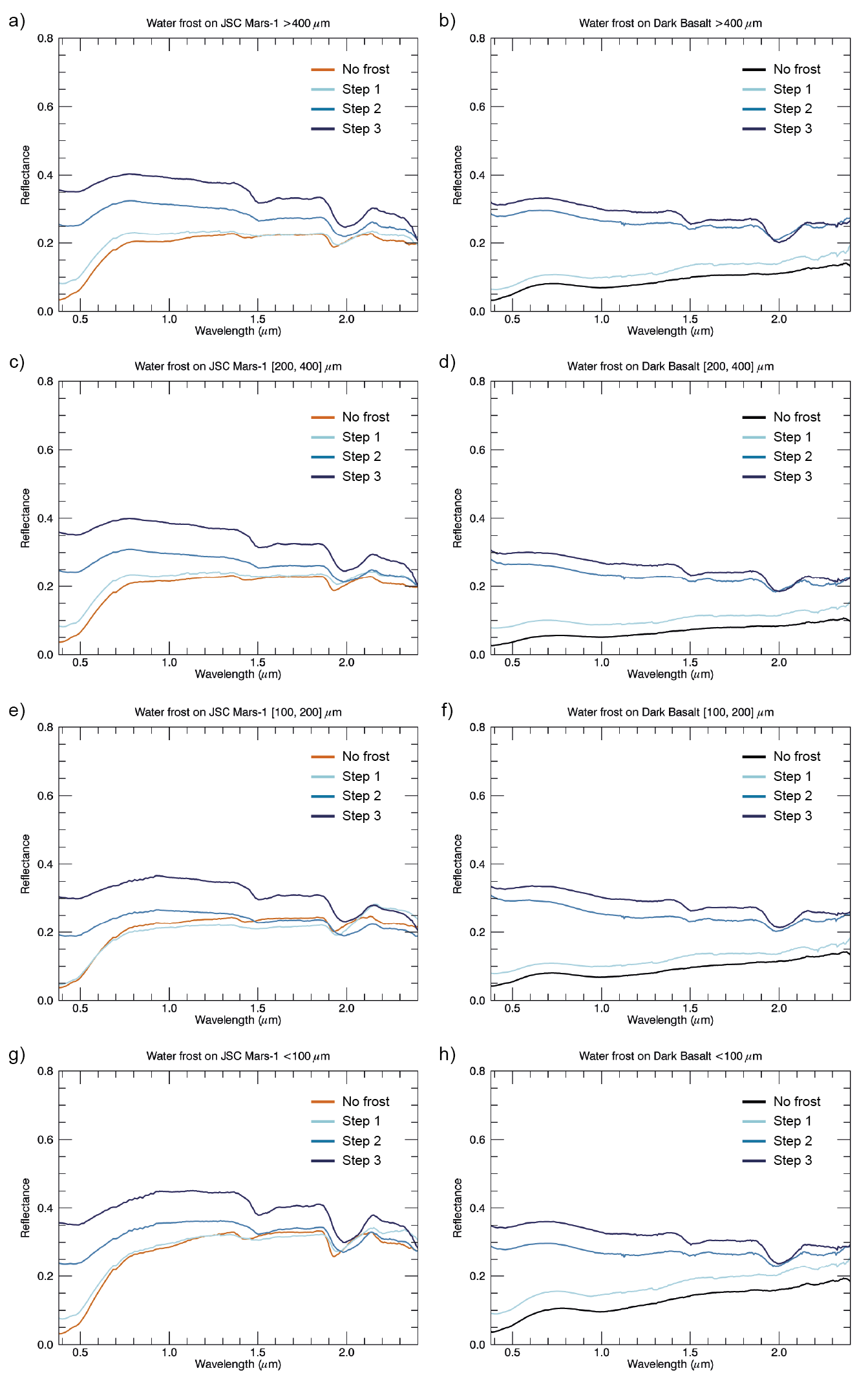}
  \caption{Reflectance spectra of JSC Mars-1 (left) and Dark Basalt (right) samples covered by water frost. Each step corresponds to 10 minutes of exposure to the atmospheric water. Each row corresponds to different size fractions of the regolith; from top to bottom, \textgreater400 \textmu m, [200,400] \textmu m, [100,200] \textmu m, and \textless100 \textmu m. }
  \label{Fig:spectra_frost}
\end{figure}

Figure \ref{Fig:spectra_frost} shows the reflectance spectra of three condensations of atmospheric water on different size-distributions of JSC Mars-1 (left) and Dark Basalt (right). While the first step of frost deposition doubles the reflectance of the Dark Basalt, it has limited impact on the spectra of the JSC Mars-1: the hydration bands slightly weaken and shift towards longer wavelengths. 

By the second step of frost deposition, both materials reflect, at the bluest wavelengths, between six and eight times more than when they were frost-free. The growth in the near-infrared reflectance is nevertheless much slower: around 2.4 \textmu m, the reflectance of the Dark Basalt has increased by about 50$\%$, while the one of the JSC Mars-1 remains roughly the same as without frost. The differences in the reflectance rise between spectral ranges results in bluer visible and near-infrared slopes for both materials. On the other hand, the last deposition of frost strengthens the absorption bands more than it increases the overall reflectance. 

\subsubsection{Intimate mixtures}
\label{S:4.1.2}
\begin{figure}
\centering\includegraphics[width=\textwidth]{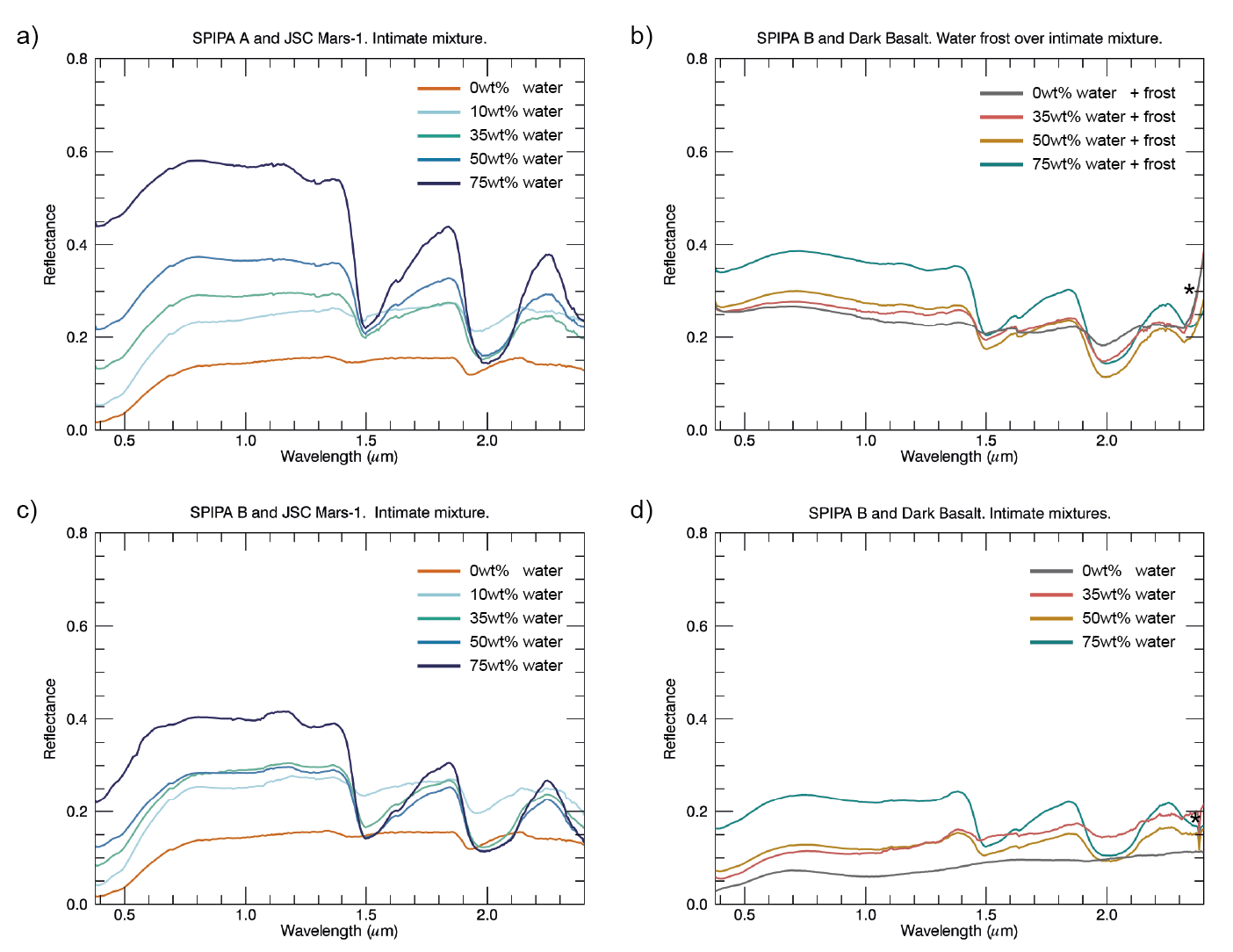}
\caption{Reflectance spectra of intimate mixtures of JSC Mars-1 (average size, 24 \textmu m) and Dark Basalt (50wt$\%$ of the particles \textgreater 109 \textmu m and 50 $\%$ \textless 109 \textmu m) with water ice (SPIPA-A: 4.5$\pm$2.5 \textmu m; SPIPA-B: 67$\pm$31 \textmu m). Different colours represent different percentages of water ice within the mixture. a) JSC Mars-1 mixed with SPIPA-A b) Dark Basalt and SPIPA-B mixtures with water frost condensed on it. The star indicates an artefact caused by the instrument c) JSC Mars-1 mixed with SPIPA-B d) Dark Basalt and SPIPA-B.}
\label{Fig:spectra_intimate}
\end{figure}

Because small particles have a greater surface-to-volume ratio than large particles, mixtures with SPIPA-A are expected to be brighter than with SPIPA-B \citep{GRL:GRL53248}. Spectra in Figures \ref{Fig:spectra_intimate}a and \ref{Fig:spectra_intimate}c show this behaviour, except for the samples with less ice. The reason for this could be the sublimation of the ice at the surface since the small SPIPA-A particles sublimate much faster than the SPIPA-B particles. 

The addition of SPIPA-A to a sample of JSC Mars-1 progressively increases the reflectance of the continuum and the H$_2$O indices. An increasing concentration of SPIPA-B also increases the water index. Nevertheless, it seems to affect the continuum of JSC Mars-1 in three phases; a first doubling of the reflectance value, followed by a stage of no change (35-50wt$\%$) and finally a new increase in the reflectance of about 50$\%$. We observe a similar sequence in dark basalt and SPIPA-B intimate mixtures (Figure \ref{Fig:spectra_intimate}d); a first increase of reflectance at 35wt$\%$ of ice is followed by a phase where the absorption bands become deeper in the near-infrared without significant changes in the visible reflectance. The sample with 75wt$\%$ of ice doubles the reflectance of that with 50wt$\%$ of ice in the visible and shows distinct absorption bands in the near-infrared.

\paragraph{Frost over an intimate mixture}
Figure \ref{Fig:spectra_intimate}b shows the reflectance spectra of frost over the intimate mix of SPIPA-B and Dark Basalt. As observed in Section \ref{S:4.1.1}, frost increases further the reflectance at shorter wavelengths, turning the slopes to blue in both the visible and near-infrared. Furthermore, the absorption bands of water deepen because of the deposition of frost. 

\subsubsection{Frozen soil}
\label{S:4.1.3}
\begin{figure}
\centering\includegraphics[width=\textwidth]{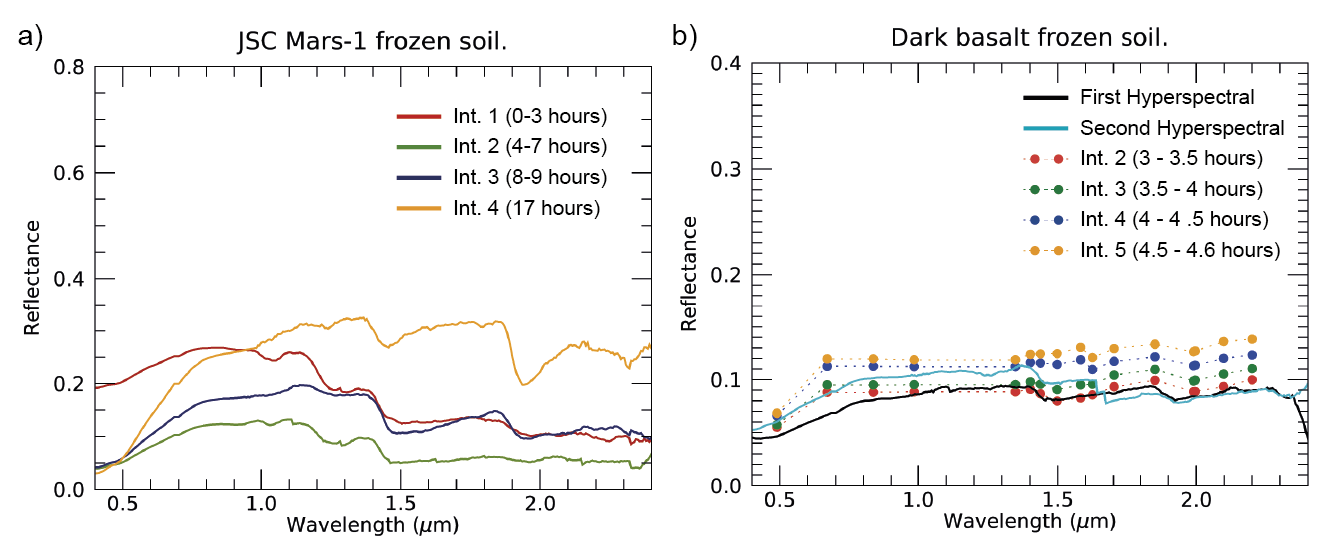}
\caption{Evolution of reflectance spectra of frozen soils of JSC Mars-1 (left) and Dark Basalt (right) through 17 and almost 5 hours of sublimation respectively. The reflectance spectra have been grouped into Intervals. In the Dark Basalt plots, the solid lines represent hyperspectral measurements, and the points represent the reflectance measured at specific wavelengths in the multispectral mode.}
\label{Fig:spectra_soils}
\end{figure}

Frozen soils prepared with JSC Mars-1 and Dark Basalt were exposed to vacuum and left to sublimate for 17 hours and almost 5 hours respectively; Figure \ref{Fig:spectra_soils} shows the evolution of their spectra during that time. To simplify the figures, we have grouped the spectra in intervals. The duration of these intervals, the pressure and temperature conditions, and the number of spectra measured per interval is specified in Figure \ref{Fig:spectra_soils} and Table \ref{Table:mixtures}.

Regardless of what the spectrum of the underlying material was, the layer of ice that covered it was thick enough to reduce its reflectance and to saturate the spectra at wavelengths longer than 1.4 \textmu m, as seen previously in \citet{Clark:1986}. For both simulants, reflectance seems to fall more in the near-infrared than in the visible. Furthermore, the absorption bands at 1.5 \textmu m and 2.0 \textmu m are saturated, leaving the near-infrared spectra featureless.

Both in the VIS and the NIR, the reflectance of the JSC Mars-1 samples decreases from the first to the second interval. We explain that by the presence of cracks within the ice, which increase the interfaces within the ice layer. Once the cracks anneal, the reflectance drops. From interval 2, the progressive sublimation of ice increases the reflectance of the samples in the NIR. In the VIS, the reflectance spectra start to adopt the shape of the JSC Mars-1 as the ice sublimates. Hence, the reflectance only increases as ice sublimates at wavelengths longer than 0.5 \textmu m. We observe in Figure \ref{Fig:spectra_soils}a that JSC Mars-1 has adsorbed water and is now highly hydrated. 

\subsection{Spectral and Colour analysis}
\label{S:4.2}
Figure \ref{Fig:criteria_water} shows the evolution of the spectral parameters defined in Section \ref{S:3.3.1} as a function of the ice content of the sample. The experiments with JSC Mars-1 are shown on the left and the ones with the Dark Basalt are shown on the right. Figures \ref{Fig:criteria_water}\textit{i} and \ref{Fig:criteria_water}\textit{ii} contain the legends applicable to each column. We do not show the frozen samples in these figures since there is not an assumed linearity in the content of water: we cannot measure the content of water per interval and the intervals do not have the same duration (see Table \ref{Table:mixtures}). 

\begin{figure}
\centering
\includegraphics[width=1.1\textwidth]{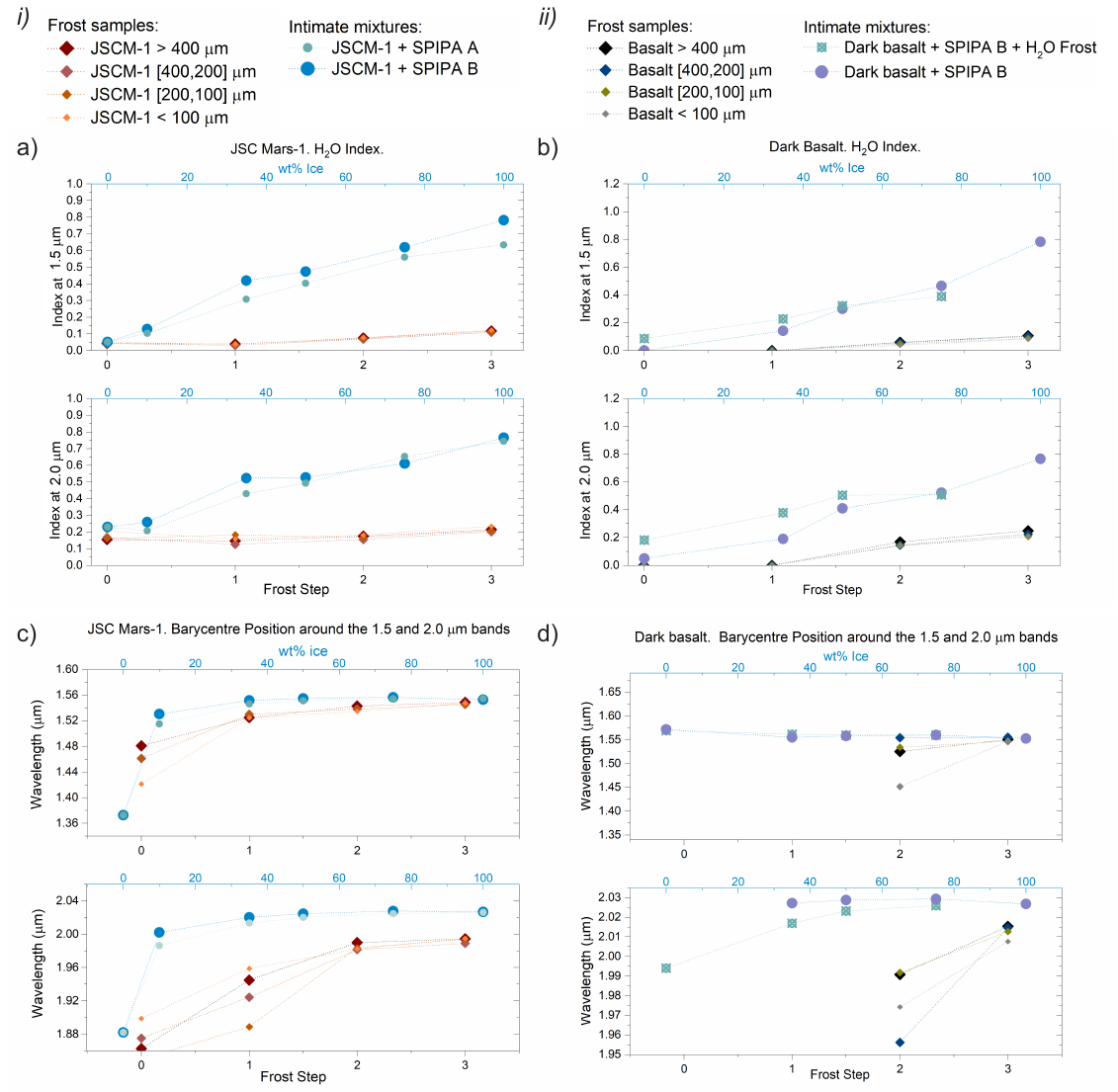}
\caption{Spectral criteria as the amount of water ice increases for JSC Mars-1 (left) and Dark Basalt (right) samples. This figure continues in Fig.\ref{Fig:criteria_water_2}. The points that correspond to frost experiments follow the X-axis at the bottom, whereas the points that correspond to intimate mixtures follow the X-axis at the top of the graphs (in blue). The top X-axis (in blue, wt $\%$ of water) and the top X-axis (in black, Frost Step) are not to be compared in terms of water content. i), ii) Legends for the JSC Mars-1 and Dark Basalt measurements respectively a), b) H$_2$O index at 1.5 \textmu m (top) and 2.0 \textmu m (bottom). c), d) Barycentre position for the band at 1.5 \textmu m (top) and 2.0 \textmu m (bottom)}
\label{Fig:criteria_water}
\end{figure}

\subsubsection{H$_2$O index}
\label{S:4.2.1}

Figures \ref{Fig:criteria_water}a and \ref{Fig:criteria_water}b show the H$_2$O indexes as a function of the amount of water. We have explained in Section \ref{S:3.3.1}, the methodology to assess the presence of a band; the lack of a point implies the lack of a band. We confirm the different impact of the intimate mixtures and the frost depositions on the H$_2$O indices; samples covered by less than 100 $\mu$m of frost show similar H$_2$O indices than intimate mixtures with a 10 wt$\%$ of water. When condensed upon the intimate mixtures of Dark Basalt and SPIPA-B, frost affects the water indexes of the samples at 2.0 \textmu m in different ways.

Coarse ice particles (SPIPA-B) intimately mixed with JSC Mars-1 produce slightly higher H$_2$O indices at 1.5 \textmu m than the fine ice particles (SPIPA-A) do. We do not observe such a trend at 2.0 \textmu m. In the experiments of frost deposition, we see no substantial influence of the particle size of the substratum in the H$_2$O indices. Only at the first frost deposition over JSC Mars-1, we measure different indices at 2.0 \textmu m. This may suggest that the different size-distributions trapped different amount of frost in the first step.

\subsubsection{Position of the barycentre}
\label{S:4.2.2}
In Figures \ref{Fig:criteria_water}c and \ref{Fig:criteria_water}d, we show the position of the barycentres for the bands around 1.5 and 2.0 \textmu m. 

The bands of the intimate mixtures with JSC Mars-1 quickly move towards the positions of the barycentre of water ice, measured for both SPIPA -A and -B at 1.55 and 2.02 \textmu m. SPIPA -A and -B seem to be equally efficient in shifting the bands, except for the sample with a 10 wt $\%$ of ice, in which the samples with SPIPA -A centre at shorter wavelengths than the ones with SPIPA-B. Frost also shifts the position of the barycentres towards long wavelengths, at different rhythms depending on the band. Hence, Step 1 of frost deposition shifts the band around 1.5 \textmu m as much as the intimate mixture with 10 wt $\%$ of ice. By the third deposition, that band is already centred at 1.55 \textmu m. The barycentre of the band around 2.0 \textmu m shifts more slowly, so that by the third step of deposition the band has shifted less than the one of any intimate mixture. For both bands, the positions we get for the JSC Mars-1 spread with the size of the dusts. The same spread is observed during the first frost deposition over JSC Mars-1.  
Intimate mixtures with Dark Basalt also align quickly around 1.55 \textmu m, with no difference if frost is present on the samples. Even if pure Dark Basalt and pure Dark Basalt with frost seem to centre around slightly longer wavelengths than pure ice (1.571 and 1.569 \textmu m respectively), the result lies in the range of our uncertainties, so no conclusions can be drawn. The band around 2.0 \textmu m, not present for Dark Basalt, appears when frost is deposited on it. As the concentration of ice within the intimate mixtures grows, the barycentres shift towards 2.02 \textmu m, more slowly if frost covers the mixtures. Regarding the frost experiments, and following the criteria explained in Section \ref{S:3.3.1}, the bands only appear on the second step of the deposition. The positions of the barycentres for Step 2 of deposition spread with the size of the dust. As observed with the JSC Mars-1, Step 3 centres the first band at 1.55 \textmu m, whereas the second band remains at shorter wavelengths than the band of the intimate mixtures.

\begin{figure}
\centering
\includegraphics[width=1.1\textwidth]{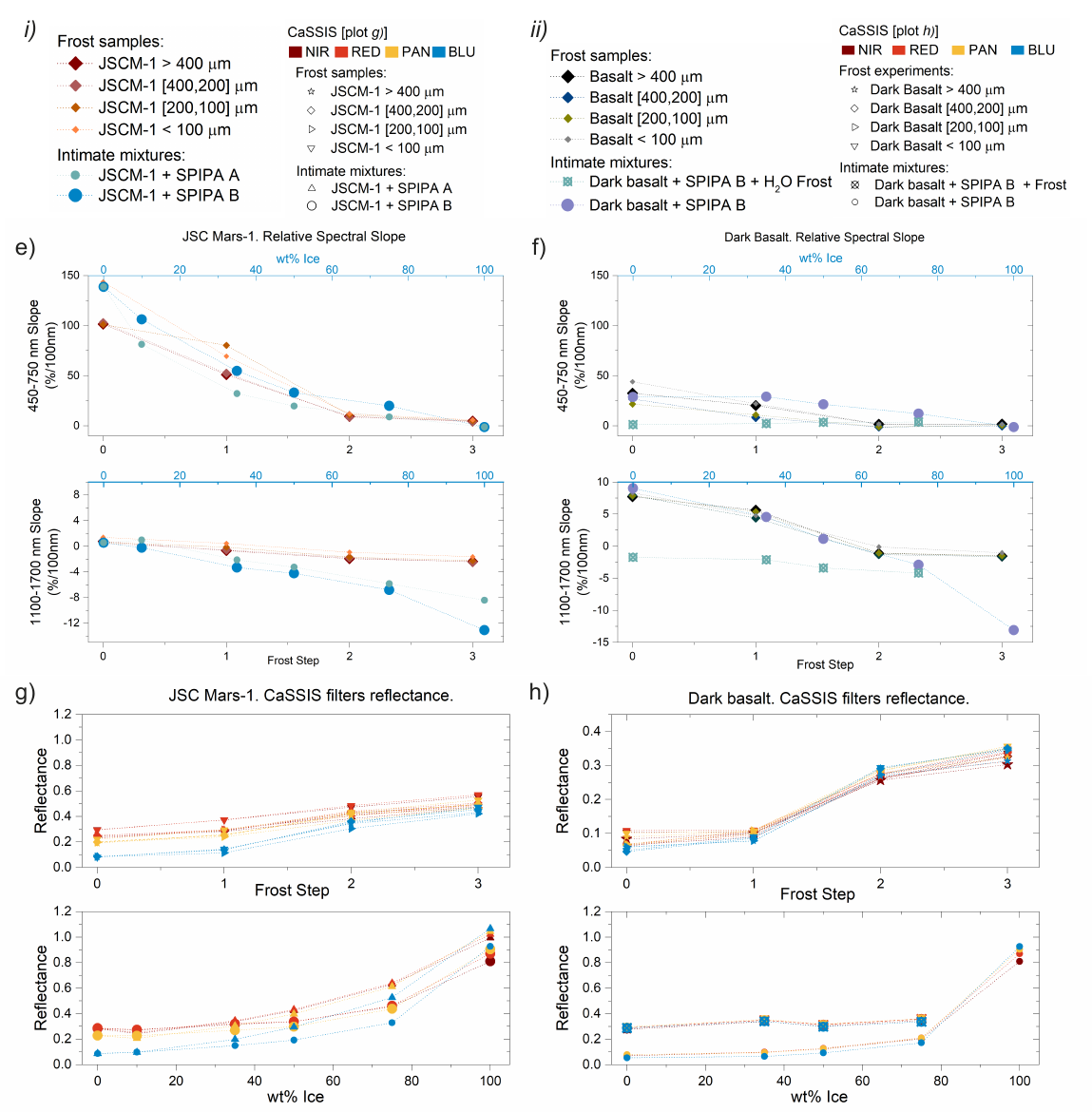}
\caption{(Continuation of Figure \ref{Fig:criteria_water}) Spectral criteria as the amount of water ice increases for JSC Mars-1 (left) and Dark Basalt (right) samples. i), ii) Legends for the JSC Mars-1 and Dark Basalt measurements, respectively e), f) VIS (top) and NIR (bottom) spectral slopes. The points that correspond to frost experiments follow the X-axis at the bottom, whereas the points that correspond to intimate mixtures follow the X-axis at the top of the graphs (in blue). Both X-axis are not to be compared in terms of water content. g), h) Reflectance measured with the CaSSIS filters for the frost experiments (top) and the intimate mixtures (bottom).}
\label{Fig:criteria_water_2}
\end{figure}

\subsubsection{Spectral Slope}
\label{S:4.2.3}
In its dry form, JSC Mars-1 shows a steep red slope in the visible and a plateau in the near-infrared (Figure \ref{Fig:criteria_water_2}e). In comparison, the slope of the dry Dark Basalt is two to six times bluer in the visible and three to seven times redder in the near-infrared, depending on the size distribution (Figure \ref{Fig:criteria_water_2}f). Nevertheless, regardless of the initial value of the visible slope, roughly 100 $\mu$m of water frost are enough to achieve similar slopes as the ones of pure particulate ice. Ice particles in intimate mixtures also reduce the initial red slopes of the materials; nevertheless, concentrations of water ice higher than 80wt$\%$ are needed to flatten the visible slope as much as tens of micrometres of frost would do. In the visible range, smaller particles flatten the slope more effectively.

In the near-infrared, all the slopes have turned to blue --- as expected from water ice. At similar concentrations of water ice, intimate mixtures look bluer with JSC Mars-1 than with Dark Basalt. At the second frost deposition, both materials already show the same spectral slope. The drop in the spectral slope of the Dark Basalt and intimate ice mixtures caused by frost confirms that, in the near-infrared, frost acts as well as an efficient slope flattener, as it does in the visible range. 

\subsubsection{Colour Analysis}
\label{S:4.2.4}
Simulating the reflectance values measured in the different filters of CaSSIS (Figures \ref{Fig:criteria_water_2}g and \ref{Fig:criteria_water_2}h) allows us to distinguish between the experiments with JSC Mars-1 and Dark Basalt. Because of the steep slope in the visible of the JSC Mars-1 analogue, we would measure different values of reflectance in each filter, whereas the reflectance of the Dark Basalt with ice would be the same in every filter. 

Through successive frost deposition steps, the reflectance of both materials rises similarly. In the case of JSC Mars-1, less than 100 $\mu$m of frost deposited over a sample would produce an increase in the reflectance two times greater in the BLU than in the RED filter. Hence, the values of reflectance in the different channels converge progressively to similar values as the condensation of frost increases. The condensation of frost on the intimate mixtures levels out the reflectance, which hampers the estimation of the amount of ice underneath. 

The curve drawn by the increasing reflectance with the growing concentration of ice differs slightly from material to material: while in the JSC Mars-1--mixtures the reflectance increases gradually from 10wt$\%$ ice content, that of the Dark Basalt-- blends only varies significantly from 75wt$\%$ of water ice. Even though pure SPIPA-A and -B are blue (i.e., the highest reflectance is measured in the BLU filter), the intimate mixtures with 75wt$\%$ of ice still show weaker values of reflectance in the BLU filter. Also for intimate mixtures of JSC Mars-1 and water ice, the filters become sensitive to the size of the ice particles as from concentrations of 35-50wt$\%$ of water ice. 

\subsection{Comparison of criteria}
\label{S:4.3}
In Figures \ref{Fig:comparison_criteria} and \ref{Fig:comparison_criteria2}, we compare, two-by-two, spectral parameters for the different mixtures. We aim to identify trends in the reflectance of the dusts when mixed in different ways with various amounts of water ice. Here, unlike in Section \ref{S:4.2}, the focus lies on the mixing style rather than on the amount of water contained in the sample; we want to identify the best spectral parameters to identify the mixing styles. We only indicate with an arrow, for orientation, the direction of increasing ice content within the sample.

The spectral parameters that we compare here provide us with a good overview of what can be done with laboratory data. For clarity, we have split the plots between JSC Mars-1 and Dark Basalt experiments, shown on the left and right column, respectively. Both Figures \ref{Fig:comparison_criteria} and \ref{Fig:comparison_criteria2} \textit{i)} show the legend to follow the plots.  

\begin{figure}
\centering\includegraphics[width=0.9\textwidth]{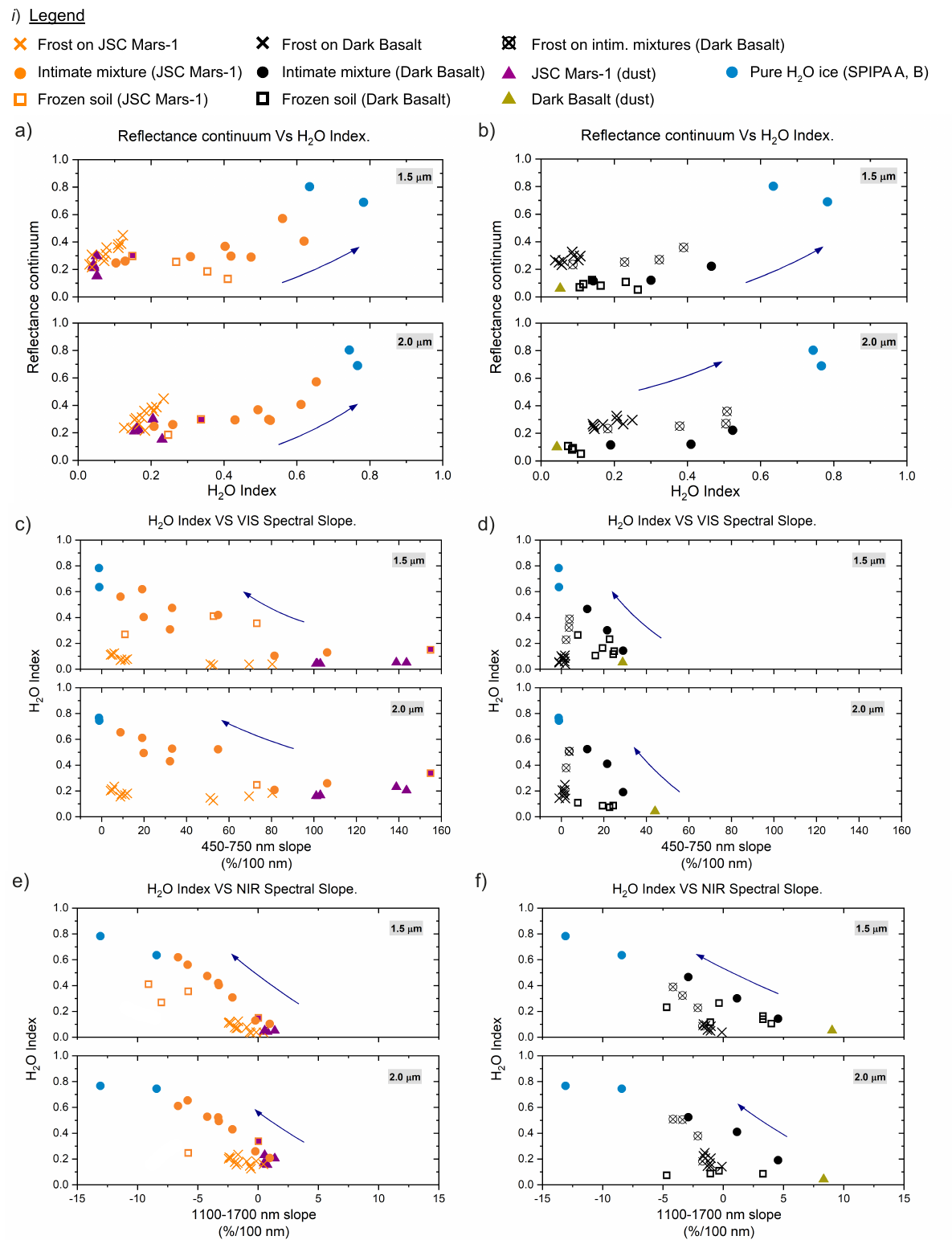}
\caption{Comparison of various spectral parameters of the laboratory samples for different ice and dust mixing modes. Experiments with JSC Mars-1 and Dark Basalt are shown on the left and right column, respectively. \textit{i)} Legend. Arrows indicate the direction of increasing content of ice in the samples. a), b) Reflectance at 1.08 \textmu m versus the H$_2$O index at 1.5 \textmu m (top) and 2.0 \textmu m (bottom) c), d) H$_2$O index at 1.5 \textmu m (top) and 2.0 \textmu m (bottom) compared to the VIS spectral slope e), f) H$_2$O index at 1.5 \textmu m (top) and 2.0 \textmu m (bottom) compared to the NIR spectral slope.}
\label{Fig:comparison_criteria}
\end{figure}

\subsubsection{Spectral parameters}
In Figures \ref{Fig:comparison_criteria}a and \ref{Fig:comparison_criteria}b, we compare the reflectance of the continuum of our samples (JSC Mars-1 in Figure \ref{Fig:comparison_criteria}a and Dark Basalt in Figure \ref{Fig:comparison_criteria}b) with the H$_2$O indices at 1.5 and 2.0 \textmu m. We observe different trends depending on the mixing modes.

Frost experiments on dusts cluster around relatively weak water indices (i.e, \textless0.2 at 1.5 \textmu m and \textless0.3 at 2.0 \textmu m) and values of reflectance ranging from 0.1 to 0.5. The growing condensation of frost on dust increases the reflectance and the water indices linearly; on hydrated minerals (JSC Mars-1), the reflectance rise is faster than that of the water index, whereas in the dry minerals (Dark Basalt) water indices increase faster than the reflectance of the continuum. 

For intimate mixtures, the reflectance of the continuum does not increase linearly with the water indices (Figures \ref{Fig:comparison_criteria}a and \ref{Fig:comparison_criteria}b) . Instead, we observe the same regimes as in the reflectance spectra (Figures \ref{Fig:spectra_intimate}a, \ref{Fig:spectra_intimate}c and \ref{Fig:spectra_intimate}d); a first one where the reflectance of the samples rises, a second one in which the water indices increase while the reflectance remains constant, and a third one where the reflectance increases exponentially with water indices. Where the transition between regimes two and three happens depends on the grain size of the ice and the albedo of the regolith. Back in Figure \ref{Fig:comparison_criteria}b, intimate mixtures with frost on them show a parallel trend to the intimate mixtures without frost, only shifted in reflectance due to the reflectance of frost. Finally, frozen soils show generally low reflectance and a wide range of water indices that can go up to 0.5 for the 1.5 \textmu m band of JSC Mars-1 frozen soils. In frozen soils experiments, the water index is not proportional to the amount of water since when a thick layer of ice covers the dust, the absorption bands saturate. In the plots of JSC Mars-1 (that is, left-hand column plots) in Figures \ref{Fig:comparison_criteria} and \ref{Fig:comparison_criteria2}, we have marked a point with an orange square filled with purple: this point represents the last measurement of the frozen soil, in which JSC Mars-1 was exposed and ice-free but, because the regolith had been in liquid water, the dust adsorbed water and is highly hydrated.  

By comparing these parameters (reflectance of the continuum and water indices), we can distinguish intimate mixtures with at least 35wt$\%$ of water ice from other mixing types. By looking at the band at 1.5 \textmu m, we can also distinguish between frozen soils and frost-covered soils.

We plot in Figures \ref{Fig:comparison_criteria}c and \ref{Fig:comparison_criteria}d the H$_2$O index as a function of the visible slope. We observe once again the flattening effect of frost: as from the second step of frost condensation, all the frost experiments gather in the region of spectral slopes weaker than 15 $\%$/100 nm. This behaviour of frost allows us to distinguish intimate mixtures with frost on them from frost-free intimate mixtures, which could not be done in Figure \ref{Fig:comparison_criteria}b. Nevertheless, it is still impossible to distinguish the smallest amounts of frost on dusts from intimate mixtures with 10wt$\%$ of water. Finally, looking at the band at 1.5 \textmu m we can distinguish frozen soils from frost experiments, while the band at 2.0 \textmu m allows us to distinguish frozen soils from intimate mixtures with more than 10wt$\%$ of water.

To distinguish between frozen soils and other samples, we plot in Figures \ref{Fig:comparison_criteria}e and \ref{Fig:comparison_criteria}f the H$_2$O indices versus the spectral slopes in the near-infrared. Compared to other experiments that contain JSC Mars-1, frozen soils show blue slopes (between -5 and -10 $\%$/100 nm) at low water indices (0.2-0.3). This clustering around blue spectral slopes and weak water indices allows us to distinguish frozen soils from the rest of experiments. Because frozen soils of dark basalt do not reach blue slopes more pronounced that -5 $\%$/100 nm, this comparison of only NIR parameters does not allow us to pinpoint those experiments. Finally, we show a comparison between the VIS and NIR spectral slopes in Figures \ref{Fig:comparison_criteria2}g and \ref{Fig:comparison_criteria2}h. This comparison is useful to identify the experiments with smallest amounts of frost from the dry powders. 

\begin{figure}
\centering\includegraphics[width=\textwidth]{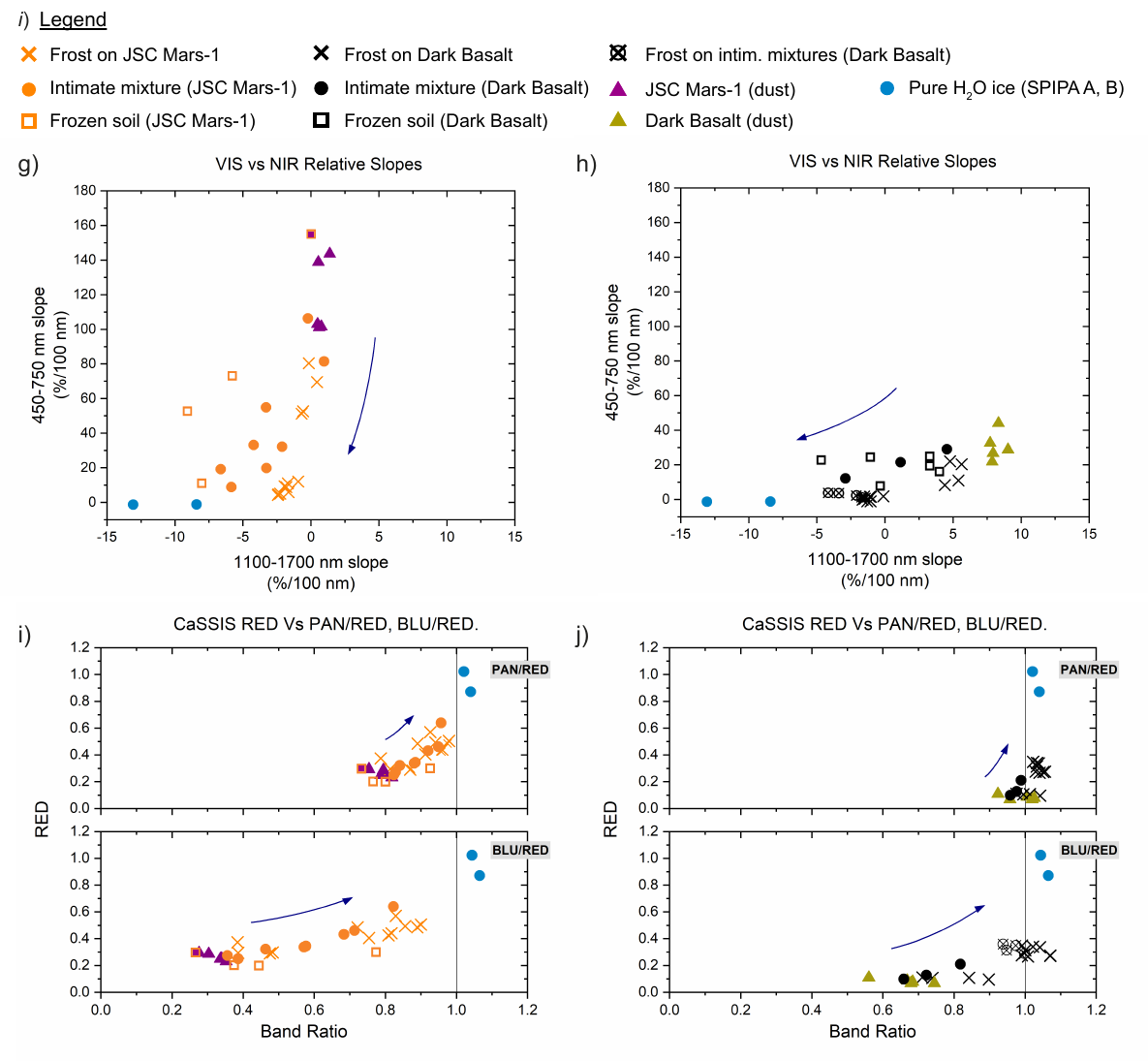}
\caption{(Continuation of Figure \ref{Fig:comparison_criteria}). Comparison of various spectral parameters of the laboratory samples for different ice and dust mixing modes. Experiments with JSC Mars-1 and Dark Basalt are shown on the left and right column, respectively. \textit{i)} Legend. Arrows indicate the direction of increasing content of ice in the samples. g),h) Comparison of the spectral slopes in the VIS and the NIR i), j) Reflectance measured in the RED channel of CaSSIS compared to the band-ratios PAN/RED (top) and BLU/RED (bottom)}
\label{Fig:comparison_criteria2}
\end{figure}

\subsubsection{Colour analysis}
The reflectance of our samples as would have been measured with CaSSIS is shown in Figures \ref{Fig:comparison_criteria2}i and \ref{Fig:comparison_criteria2}j.  In this case, we compare the signal measured in the RED filter with two band ratios: PAN/RED and BLU/RED. Ratios lower than one correspond to red slopes between the centre of the considered bands, ratios higher than one indicate blue slopes and ratios around one flat slope. 

In this plot, mixtures with JSC Mars-1 and mixtures bearing Dark Basalt are easy to separate, since they have different spectral slopes in the visible. We also see that the BLU/RED plot looks like a stretched version of the PAN/RED one, which indicates a linear correlation between the reflectance measured in the BLU and PAN filters. 

In the BLU/RED plot, we can see that CaSSIS is sensitive enough to detect the lowest amounts of frost. The arrow indicates the direction of increasing abundance of ice; we are therefore able to distinguish the first step of frost deposition from the dry samples and from the other deposition steps, which cluster between 0.7 and 0.9 BLU/RED ratio. However, frost experiments cannot be discerned from intimate mixtures only with CaSSIS.

\section{Discussion.}
\label{S:5}

\subsection{Effect of the mixing mode in the reflectance spectra.}
So far, we have seen the reflectance of various ice and dust mixtures. We have also observed how the way in which end members mix affects some specific spectral parameters. Now, we discuss further this effects and how to take advantage from them to identify different mixing modes from their reflectance spectra.

\subsubsection{Frost}
We have identified three spectral regimes as frost condenses on a surface. In the first regime, frost covers almost homogeneously the substrate with fine particles (on the order of microns). Hence, the probability for a photon to encounter a grain of ice is large, and because of the great surface-to-volume-ratio of tiny particles, the reflectance of the surface rises slightly. This is what we see in the Step 1 of our frost depositions. Since the transition from Rayleigh to a more efficient particulate scattering occurs more rapidly at short wavelengths \citep{Kieffer:1970}, at this stage the reflectance increases faster at short wavelengths. We observe this behaviour in the transition between Steps 1 and 2. The deepening of the absorption bands of the spectrum at this stage of frost deposition is limited, because of the small size of the particles of frost. 

In a second stage, ice grows thick. This does not significantly increase the probabilities for photons to encounter ice, but the path length of photons inside ice does increase. The second stage is, therefore, characterised by a deepening of the absorption bands rather than an increase in reflectance, as see in our third step of deposition. Finally, at some point, the layer of ice becomes optically thick enough to cover the spectral signature of the material underneath. However, we have not explored this stage in our experiments, and such amount of frost is not expected in frost depositions over the Martian nights. The frost that condenses at the beginning of fall and stays over the winter, a.k.a seasonal frost, is expected to sinter and grow into large crystals that do not show the same spectral properties as the light frosts that we observe in here. It should be debated whether the \textit{seasonal frost} should be referred to as \textit{frost}.

\subsubsection{Intimate mixtures}
Previous laboratory work has shown that the reflectance of water ice and dust intimate mixtures does not increase linearly with the amount of water ice \citep{GRL:GRL53248, gyalay2019}. Instead, it is dominated by the most absorptive component until the least absorptive material exceeds a threshold relative abundance, which is determined by the grain-size and albedo of the components. If we take the example of our ice and JSC Mars-1 intimate mixtures, JSC Mars-1 is the absorptive material except around 1.5 and 2.0 microns, where water ice is more absorptive than JSC Mars-1. As the amount of water ice increases from 0wt$\%$ to 100wt$\%$, we identify three phases. 

The first one can be seen in the mixtures with JSC Mars-1 and 10wt$\%$ of water ice, where the reflectance of the sample doubles since ice scatters the incoming photons more efficiently than JSC Mars-1 does. In a second phase, the spectra are characterised by different behaviour in the visible and the near-infrared, as can be seen in the intimate mixtures with 35-50wt$\%$. When we increase the amount of ice in one of our mixtures, the relative abundance of absorptive particles decreases in the visible but increases at some wavelengths in the NIR (around the absorption bands of water). Because the most absorptive component dominates the reflectance of an intimate mixture, the spectrum of JSC Mars-1 dominates in the VIS region whereas the spectrum of water ice dominates in the NIR regions: the reflectance in the VIS stays constant or rises smoothly (depending on the size of the ice) whereas the absorption bands deepen steadily. In a final, third step, the abundance of ice is enough to dominate both the VIS and NIR spectra. 

\subsubsection{Frozen soils}
Layers of ice over soils are challenging to detect since water ice is almost a continuum medium, and therefore frozen soils can be transparent in the visible range and saturated---and therefore featureless---in the near-infrared. When cracks appear in the ice, they create interfaces within the ice layer that increase the reflectance. As the layer of ice on top of the soil gets thin, the spectrum of the soil is revealed; if this soil is bright or has characteristic features, it will be detectable. Otherwise, as in the case of the dark basalt, frozen soils---both as "ice-covered soil" and "dust included in a matrix of ice"--- are nearly indistinguishable from the dry soil.

\subsection{Discriminating spectral/colour criteria to identify different ice/dust mixing modes.}
\label{S:5.4}
We propose now the following scenario: if we were given a random point from Figures \ref{Fig:comparison_criteria} and \ref{Fig:comparison_criteria2}, how could we identify to what type of mixture it corresponds? 

First, we could take advantage of the saturated near-infrared spectra of frozen soils to identify them. Indeed, frozen soils show blue slopes akin to the ones of fine-grained ice but their H$_2$O indexes up to three and even eight times weaker (at 1.5 \textmu m and 2.0 \textmu m respectively). Furthermore, unlike intimate mixtures, frozen soils conserve the water absorption band at 1.03 \textmu m when they have a thick layer of water ice on top, which differentiates them from the other mixing modes.

In order to identify frost deposits, it is a good strategy to compare the water indexes with the spectral slopes; frost neutralises the slopes without deepening the absorption features, while other forms of water ice also deepen the absorption bands (Figures \ref{Fig:comparison_criteria}c, \ref{Fig:comparison_criteria}d). Frost is easier to detect when the substrate has a steep slope, as seen in experiments with JSC Mars-1. Intimate mixtures with frost on them can also be identified in this comparison. Early stages of frost deposition can be differentiated from dry soils by comparing the visible and the near-infrared spectral slopes.

It has been shown that it is not always possible to detect water ice intimately mixed with soil in the visible range at specific geometries \citep{GRL:GRL53248}. Instead, comparing that reflectance to the H$_2$O index depths of water represents a good alternative to identify intimate mixtures of ice and dust. This comparison does not allow one to rule out the presence of frost over icy intimate mixtures; they appear like intimate mixtures of fine-ice particles (Figure \ref{Fig:comparison_criteria}a). 

For our laboratory samples, we could guarantee the identification of frost-free intimate mixtures by process of elimination: we could first rule out frozen soils and frost by the methods explained above, leaving the intimate mixtures as the remaining option. In natural surfaces, this process is not that straightforward, because other dust and ice mixtures, such as intra-mixtures (i.e., when dust grains are included in the ice grains), can have similar reflectance spectra to those of the intimate mixtures. Other studies have shown that inter- and intra-mixtures could be distinguished in the visible range \citep{Jost:2017} and, if the ice at the surface sublimates, by the evolution of their bands \citep{Jost:2017} or the properties of the sublimation residues \citep{Poch2016154}.

\subsection{Frost and intimate mixtures VS hydrated minerals}
Identifying from remote sensing whether an absorption band is caused by structural water or the deposition of frost (or even low contents of intimately mixed water ice) on the surface remains a challenge. With an \textit{a priori} knowledge of the ice-free spectrum of a hydrated material, some of the comparisons between parameters allow one to identify early stages of frost deposition (e.g., Figures \ref{Fig:comparison_criteria}b, \ref{Fig:comparison_criteria}d), but none of them is conclusive without the prior knowledge of the hydration state. 

The study of the position of the barycentres provides us with a good way of differentiating hydration from frost (see \ref{Fig:criteria_water}c). Nevertheless, the way in which we compute the barycentre in this paper requires a high sampling and spectral resolution. Instruments like OMEGA or CRISM are needed to conduct such analysis. 

\subsection{On the pertinence of the laboratory samples}
Here, we discuss the extent to which the samples that we have produced in the laboratory match the features of Martian ices. 

The nature of Martian frost is not yet well known. It follows that a direct comparison between Earth- and Mars-grown frosts can, at the moment, be only speculative. However, there are some points that we can already take into consideration. \citet{Kieffer_1968} stated that, for a process limited by diffusion, the size of the frost crystals is proportional to the relative concentration of the condensible gas. Hence, we can expect the Martian frost to be finer than the ones grown in the terrestrial atmosphere for a same period of deposition. Furthermore, it has to be taken into account that in the case of our laboratory experiments, the cooling comes from beneath the sample, and therefore the dust is colder than the atmosphere. This is not always true on Mars, where the atmosphere is in general colder than the regolith. If the atmosphere is cooler that the substrate, the grains of frost tend to grow tall towards the cold, irradiating the heat into space and cooling down through radiative cooling. If the substrate is colder than the atmosphere, as it is in our laboratory setup, the grains of frost cool down by conducting the heat into the substrate: since the efficiency of this cooling depends on the thermal contact of the frost grains with the substrate, the grains of frost will tend to grow larger than the ones that cool down radiatively \citep{Kieffer_1968}. Considering these points, one could expect the Martian grains of frost to be finer than the terrestrial frosts and to gather into filament-like structures. Would this be the case, Martian frosts would scatter light more efficiently than terrestrial ones, and the observed differential spectral behaviour between the VIS the NIR would be even stronger.  

Regarding the frozen soils, we do not expect to find on Mars soils produced by the process used to create the laboratory analogues. Nevertheless, these samples might be good analogues for packed snow (or ice) with dust grains in it, as we can find in the water ice sheets found under the Martian surface that sublimate when exposed to the atmosphere \citep{Dundas:2018}, or the exposed dusty ice on the polar deposits. In both these examples, the snow has likely metamorphosed into firn or ice (as it happens on terrestrial ice sheets \citep{Blunier:2000}, and therefore the path of the photons within the ice is longer than within the SPIPA-A and -B ice grains. That is precisely the interest of the frozen soil samples: unlike the frosts and intimate mixtures, the ice is not a particulate medium but a continuum, which, for the mentioned examples, constitutes a better approximation than the rest of the samples. We also note that our frozen soils analogues resemble more closely the ice-dust mixtures from \citet{Clark:1981} and \citet{Clark_1984} than our own intimate particulate mixtures.
\include{Case_Mars}
\section{Conclusion and perspectives}
\label{S:7}
We have conducted a series of laboratory experiments to measure the VIS and NIR reflectance of various water ice and regolith mixtures. The dry members of the mixtures are the Martian regolith analogue JSC Mars-1 and a more pristine dark basalt to simulate the observed red and black terrains from Mars. We have performed spectral and colour analysis on \begin{inparaenum}[1)]\item cold-trapped atmospheric water onto regolith to simulate frost deposition, \item intimately mixed dust and ice, and \item frozen, water-saturated regolith samples.\end{inparaenum} 

The main results are summarised below:

\begin{itemize}
\item We have identified three spectral regimes as frost grows on a dusty surface. There is an initial increase in the reflectance preferentially at short wavelengths; second, an increase in reflectance accompanied by a deepening of the absorption bands and finally, a masking of the surface underneath.

\item We have identified three spectral regimes as the abundance of water ice in an ice/dust intimate mixture grows. These three regimes are the result of a competition between the single scattering albedo of the materials, their relative abundance and the size of their particles. In the first regime, small amounts of water ice increase the reflectance of the mixture. In the second regime, the spectrum of the absorptive material dominates the reflectance spectrum of the mixture; there is a modest increase of reflectance in the non-absorptive spectral range of water and the absorption bands deepen proportionally to the abundance of water. In the third regime, there is enough water ice so that its reflectance spectrum dominates the spectrum of the mixture. 

\item Frozen soils are challenging to detect since the formation of a layer of water ice on top of the regolith can drop the reflectance of the ground if the ice is translucent (i.e., it does not have cracks) and if the reflectance is measured outside specular geometry. 

\item We have shown how a strategic comparison of spectral and colour criteria provides crucial information about the mixing mode of ice and dust.

\item We have compared a set of parameters in the visible and near-infrared spectral ranges. This is key to making our data useful for a maximum of Martian imagers since some of them observe only in the VIS or in the NIR, or in cases where failures cripple the capability of instruments to image the full spectral range. 

\item To differentiate intimate mixtures from frost experiments, the best is to compare VIS and NIR parameters. Intimate mixtures will, as a general rule, have more impact in the NIR, whereas frost will affect the VIS rapidly. To identify frost on an intimate mixture ---even in layers of tens of micrometres, it is useful to compare their water index to their visible spectral slope. 

\item The study of the position of the barycentres allows us to differentiate hydrated minerals from early stages of frost deposition on dust.
\end{itemize}

The future steps of our work are to study the effect of a Martian-like atmosphere in the reflectance and detectability of our samples. The first attempts to compare our laboratory data to Martian surfaces have been presented in planetary conferences and have given promising results \citep{YoldiLPSC:2018, YoldiEPSC:2018}. In the companion paper, we have added CO$_2$ ice to our experiments and conducted a similar spectral and colour analysis.

\vspace{50px}
\textbf{Data availability}
All the reflectance spectra presented here are available on the \href{https://wiki.sshade.eu/sshade/databases/bypass}{Bern icY Planetary Analogues Solid Spectroscopy (BYPASS)}  database, hosted on the \href{https://wiki.sshade.eu/start}{Solid Spectroscopy Hosting Architecture of Databases and Expertise (SSHADE)} .

\vspace{50px}

\textbf{Acknowledgements}
The authors want to thank Daniele Piazza and the Workshop team of the University of Bern for helping in the design and construction of the frost sample-holder. The authors also want to thank Martin Grosjean and Tobias Schneider from the Oeschger Centre for Climate Change Research and Institute of Geography, University of Bern, for allowing us to use their SEM. Z. Yoldi wants to thank Panagiotis Theologou for his help formatting Table \ref{Table:mixtures}. This work has been developed in the framework of the National Center for Competence in Research PlanetS funded by the Swiss National Science Foundation (SNSF). Olivier Poch acknowledges funding from CNES and European Research Council under grant SOLARYS (77169). The authors want to thank Ken Herkenhoff, Vincent Chevrier and the anonymous reviewers for their constructive and helpful inputs.



%% References


%%\bibliographystyle{model1-num-names}
\bibliography{sample.bib}






% \end{thebibliography}

\end{document}